\PassOptionsToPackage{unicode}{hyperref}
\PassOptionsToPackage{hyphens}{url}
\PassOptionsToPackage{dvipsnames,svgnames,x11names}{xcolor}
\documentclass[
  astrosymb, twocolumn, tighten, twocolappendix]{aastex631}
\usepackage{amsmath,amssymb}
\usepackage{lmodern}
\usepackage{iftex}
\ifPDFTeX
  \usepackage[T1]{fontenc}
  \usepackage[utf8]{inputenc}
  \usepackage{textcomp} % provide euro and other symbols
\else % if luatex or xetex
  \usepackage{unicode-math}
  \defaultfontfeatures{Scale=MatchLowercase}
  \defaultfontfeatures[\rmfamily]{Ligatures=TeX,Scale=1}
\fi
\IfFileExists{upquote.sty}{\usepackage{upquote}}{}
\IfFileExists{microtype.sty}{% use microtype if available
  \usepackage[]{microtype}
  \UseMicrotypeSet[protrusion]{basicmath} % disable protrusion for tt fonts
}{}
\makeatletter
\@ifundefined{KOMAClassName}{% if non-KOMA class
  \IfFileExists{parskip.sty}{%
    \usepackage{parskip}
  }{% else
    \setlength{\parindent}{0pt}
    \setlength{\parskip}{6pt plus 2pt minus 1pt}}
}{% if KOMA class
  \KOMAoptions{parskip=half}}
\makeatother
\usepackage{xcolor}
\usepackage{graphicx}
\makeatletter
\def\maxwidth{\ifdim\Gin@nat@width>\linewidth\linewidth\else\Gin@nat@width\fi}
\def\maxheight{\ifdim\Gin@nat@height>\textheight\textheight\else\Gin@nat@height\fi}
\makeatother
\setkeys{Gin}{width=\maxwidth,height=\maxheight,keepaspectratio}
\makeatletter
\def\fps@figure{htbp}
\makeatother
\providecommand{\tightlist}{%
  \setlength{\itemsep}{0pt}\setlength{\parskip}{0pt}}
\usepackage[capitalize]{cleveref}
\usepackage{CJKutf8}
\usepackage{newunicodechar}

\newcommand{\Dnu}{\ensuremath{\Delta\nu}}
\newcommand{\numax}{\ensuremath{{\nu_\text{max}}}}

\newcommand\mesa{{\textsc{mesa}}}
\newcommand\gyre{{\textsc{gyre}}}

\defcitealias{ong_surface_2021}{OBR21}
\newcommand\obr{{\citetalias{ong_surface_2021}}}

\newcommand{\chinesename}{{\begin{CJK}{UTF8}{gbsn}(王加冕)\end{CJK}}}

\DeclareRobustCommand{\okina}{%
  \raisebox{\dimexpr\fontcharht\font`A-\height}{%
    \scalebox{0.8}{`}%
  }%
}
\newunicodechar{ʻ}{\okina}

\newcommand{\annotate}[2]{\begin{tikzpicture}
    \node[anchor=south west,inner sep=0,align=center] (image) at (0,0) {
    #1
    };
    \begin{scope}[x={(image.south east)},y={(image.north west)}]
    #2
    \end{scope}
\end{tikzpicture}}

\graphicspath{{./},{./figures/}}

\renewcommand{\edit}[2]{{\ifnum#1<10%
#2%
\else%
\textbf{#2}%
\fi}}
\usepackage{mathptmx,txfonts,tikz,bm,braket}
\ifLuaTeX
  \usepackage{selnolig}  % disable illegal ligatures
\fi
\usepackage[]{natbib}
\IfFileExists{bookmark.sty}{\usepackage{bookmark}}{\usepackage{hyperref}}
\IfFileExists{xurl.sty}{\usepackage{xurl}}{} % add URL line breaks if available
\hypersetup{
  colorlinks=true,
  linkcolor={Maroon},
  filecolor={Maroon},
  citecolor={xlinkcolor},
  urlcolor={xlinkcolor},
  pdfcreator={LaTeX via pandoc}}

\date{\today}

\begin{document}

\shorttitle{Surface Term Corrections for Inversion Kernels}
\title{Red Giant Rotational Inversion Kernels Need Nonlinear Surface Corrections}
%!TEX root=kernels.tex
% ApJ style

\correspondingauthor{Joel Ong}
\email{joelong@hawaii.edu}
\author[0000-0001-7664-648X]{J. M. Joel Ong \chinesename}
\altaffiliation{NASA Hubble Fellow}
\affiliation{Institute for Astronomy, University of Hawaiʻi, 2680 Woodlawn Drive, Honolulu, HI 96822, USA}

\received{October 17, 2023}
\revised{November 11, 2023}
\accepted{November 13, 2023}
\submitjournal{\apj}
\shortauthors{Ong}
\def\sectionautorefname{Section}
\def\subsectionautorefname{Section}
\def\subsubsectionautorefname{Section}

% MNRAS Style

% \author[J. M. J. Ong and S. Basu]{
% J. M. Joel Ong,$^{1}$\thanks{joel.ong@yale.edu}
% Sarbani Basu$^{1}$
% \\$^{1}$Department of Astronomy, Yale University, 52 Hillhouse Ave., New Haven, CT 06511, USA}
\begin{abstract}
Asteroseismology is our only means of measuring stellar rotation in their interiors, rather than at their surfaces. Some techniques for measurements of this kind --- "rotational inversions" --- require the shapes of linear response kernels computed from reference stellar models to be representative of those in the stars they are intended to match. This is not the case in evolved stars exhibiting gravitoacoustic mixed modes: we show that the action of the asteroseismic surface term --- systematic errors in the modelling of near-surface layers --- changes the shapes of their inversion kernels. Corrections for the surface term are not ordinarily considered necessary for rotational inversions. We show how this may have caused previous estimates of red-giant envelope rotation rates from mixed-mode asteroseismic inversions to have been unintentionally contaminated by core rotation as a result, with errors comparable to the entire reported estimates. We derive a mitigation procedure for this hitherto unaccounted systematic error, and demonstrate its viability and effectiveness. We recommend this mitigation be applied when revising existing rotational inversions. Finally, we discuss both the prospects for applying such mitigation to the harder problem of inversions for stellar structure (rather than rotation), as well as the broader implications of this systematic error with regards to the longstanding problem of internal angular momentum transport.
\end{abstract}

\hypertarget{introduction}{%
\section{Introduction}\label{introduction}}

Almost all astronomical measurements of stellar properties are made only of their surfaces. Asteroseismology --- the analysis and interpretation of stellar oscillations --- is one of the very few ways we probe their interiors directly, through the use of so-called inversion techniques. In this paper, we focus on the rotational inverse problem \citep[e.g.][]{schunker_inversion_2016a, schunker_inversion_2016b}, whose application to red giant solar-like oscillators \citep[including but not limited to e.g][]{deheuvels_seismic_2012, deheuvels_seismic_2014, deheuvels_seismic_2015, dimauro_internal_2016, fellay_asteroseismology_2021, pijpers_asteroseismogyrometry_2021, wilson_constraining_2023} is a key lynchpin in our understanding of how angular momentum is internally transported and redistributed over the course of stellar evolution.

The nonrotating angular frequencies \(\omega_i\) of these oscillations, each associated with eigenfunctions \(\vec{\xi_i}\), are the normal-mode solutions to a Hermitian eigenvalue problem determined by the stellar structure \citep[e.g.][]{unno_nonradial_1989, gough_linear_1993}, expressed in terms of the wave operator \(\mathcal{L}\) as \(\mathcal{L}[\vec{\xi_i}] + \omega_i^2 \rho \vec{\xi_i} = 0\). The response of these frequencies to small perturbations to the wave operator, \(\mathcal{L} \mapsto \mathcal{L} + \epsilon \mathcal{F}\), may be described by Rayleigh-Schrödinger perturbation theory through expansions of the form \citep[e.g.][]{landau_quantum_1965}
\begin{equation}
    -\delta\omega_i^2 = \epsilon \left<\vec{\xi_i}, \mathcal{F}\vec{\xi_i}\right> + \epsilon^2 \sum_{j\ne i} {\left|\left<\vec{\xi_i}, \mathcal{F}\vec{\xi_j}\right>\right|^2 \over \omega_j^2 - \omega_i^2} + \mathcal{O}\left(\epsilon^3\right),\label{eq:expand}
\end{equation}
under the convention that eigenfunctions are orthonormal with respect to the inner product: \(\left<\vec{\xi_i}, \vec{\xi_j}\right> = \int \vec{\xi_i} \cdot \vec{\xi_j}\ \rho\ \mathrm d^3 r = \delta_{ij}\).
For rotation rates \(\Omega\) much lower than the critical frequency \(\omega_0 \propto \sqrt{GM/R^3}\), individual modes of degree \(\ell\) are split into multiplets of \(2\ell+1\) components each \citep[e.g.][]{gough_rotation_1990}, with a multiplet splitting given to first order in \cref{eq:expand} as
\begin{equation}
    \delta\omega_i =  \beta_i \int \Omega(r) \cdot K_{i}(r)\ \mathrm d r, \label{eq:perturb}
\end{equation}
ignoring latitudinal differential rotation. Here \(K_{i}\) is a sensitivity kernel (defined to be of unit integral, with the overall sensitivity instead specified by the normalisation factor \(\beta_i\)) for the \(i\)\textsuperscript{th} mode. These integrals against sensitivity kernels in \cref{eq:perturb} are the diagonal matrix elements \(R_{ii}\) of the angular momentum operator \(\mathcal{R}\) \citep[e.g.][]{lyndenbell_stability_1967}, whose off-diagonal matrix elements also give rise to off-diagonal kernels by way of some quadratic form \(R\) acting on the mode eigenfunctions, defined in the sense that
\begin{equation}
    \int \beta_{ij} K_{ij}(r)\ \Omega(r) \mathrm d r = R_{ij} = \left<\vec {\xi_i}, \mathcal{R}\vec{\xi_j}\right> = \int R\left[\vec {\xi_i}, \vec{\xi_j}\right] \Omega(r) \ \mathrm d r.\label{eq:bilinear}
\end{equation}
Such off-diagonal terms can be seen to enter \cref{eq:expand} only at second and higher order.

In principle, this construction describes how the splittings of an actual star would relate to its (perfectly known) interior structure. However, the inversion kernels of an imperfect model may resemble those of the actual star, as long as these imperfections are small, in which case these integral expressions may also approximately apply to the kernels of such a reference model instead. The response of the mode frequencies to such small structural deviations may also take similar form to \cref{eq:perturb}, integrating against sensitivity kernels only once over the stellar structure \citep[e.g.][]{kosovichev_inversion_1999}. This is a necessary condition for the notional ``true'' inversion kernels of the actual star to be well-described by those of the reference model used to approximate them.

One class of these rotational inversion techniques --- that of optimal local averages \citep[OLA: e.g.][]{backus_resolving_1968} --- operates on the principle that, given a sufficiently large number of observed modes, one may choose some linear combination of these sensitivity kernels, \(\sum_i c_i \beta_i K_{i}\), so as to suppress sensitivity except at a predetermined localisation radius \(r_0\) (e.g.~so that the summed kernel approximates a Gaussian of unit integral at that location). The corresponding linear combination of rotational splittings, \(\sum_i c_i \delta\omega_i\), then yields an estimate of \(\Omega(r_0)\), and thus permits \(\Omega(r)\) per se to be inferred \citep[by varying $r_0$ repeatedly; e.g.][]{jcd_differential_1996}. Operationally, then, these sensitivity kernels are found from stellar models constructed to closely match the observed properties of oscillating stars, with the inputs \(\delta\omega_i\) to the inversion problem being the observed rotational multiplet widths. Similar principles underlie inversions for stellar structure \citep[e.g.][]{basu_fresh_2009}, where the inputs are given by differences between the observed and model mode frequencies. For this purpose, the actual model mode frequencies are never directly used for this purpose --- systematic error corrections are always applied, such as for the asteroseismic ``surface term'' (correcting modelling error in the near-surface layers).

Many variations on the OLA inversion procedure exist in the literature \citep[e.g.][]{gough_inverting_1985, pijpers_sola_1994, rabellosoares_parameters_1999}, with different ways of choosing coefficients \(\{c_i\}\) (by optimising different merit/penalty functions) to yield summed kernels with various desirable properties. These all operate by penalising sensitivity of the summed kernel away from \(r_0\) in different ways, while also, for example, minimising the expected observational error of \(\sum_i c_i \delta \omega_i\), or maintaining a summed kernel of unit integral over the whole star. At their heart, they all nonetheless rely on the same fundamental assumption of linearity. This is also true of the other class of inversion techniques --- the regularised least-squares method \citep[RLS, e.g.][]{jcd_comparison_1990}, which fits constrained parameterisations of \(\Omega(r)\) by forward-modelling multiplet splittings through \cref{eq:perturb} --- also commonly applied in the literature.

While these linear inversion procedures require first-order accuracy for all modes used, the radius of convergence for perturbation theory varies from mode to mode. On this principle, \citet[hereafter \obr]{ong_surface_2021} derive a nonlinearity threshhold for evolved solar-like oscillators like sub- and red giants. Should the surface term exceed it, such first-order perturbation analysis would then poorly describe their dipole modes of mixed gravitoacoustic character. This surface term acts as a further structural perturbation to a reference model, whose action may pass the pure p-modes through a forest of avoided crossings (e.g.~their fig.~11). Generically, such avoided crossings signify mode frequencies and eigenfunctions responding nonlinearly to structural or dynamical perturbations. \citet{vanlaer_feasibility_2023} recently find that their presence in g-mode pulsators accompanies modifications to the shapes of the inversion kernels themselves, rendering structure inversions in those stars infeasible.

This has historically not been a concern in helioseismology, in which astronomical context these techniques were first applied, nor for their application to main-sequence p-mode oscillators: the size of the surface term in those stars is small relative to the nonlinearity criterion of \obr. In these stars, the surface term does not substantially modify the splitting widths given by \cref{eq:perturb}, and thus is not thought to be relevant for rotational inversions. In any case, even where applied, existing prescriptions for surface-term corrections also operate only on mode frequencies, leaving eigenfunctions and inversion kernels unmodified. Current applications of rotational inversions to evolved sub- and red giants have inherited these helioseismic techniques with few modifications. \edit1{While alternative, nonlinear, helioseismic inversion techniques exist (e.g. \citealt{marchenkov_nonlinear_2000}), these so far rely on specific features of analytic approximations to p-modes \citep{roxburgh_asymptotic_1996}, and are thus not yet suitable for use on mixed or g-modes.} However, \citet{li_surface_2023} report that typical sizes of the asteroseismic surface effect, as calibrated on Kepler red giants, are indeed large enough to advance modes along avoided crossings. Therefore, as this fundamental assumption of linearity no longer applies, the results of \obr~immediately imply that the surface term changes the shape of inversion kernels in red giants too. Similar interference to \citet{vanlaer_feasibility_2023} must thus afflict rotational inversions using these mixed modes.

We now examine how this may affect ongoing attempts at isolating seismic envelope rotation rates from red-giant dipole mixed modes. We moreover describe a well-defined mitigation procedure for the rotational inversion problem in particular. Finally, we conclude with some discussion regarding generalisations for the structural inversion problem, and broader implications.

\hypertarget{case-study-envelope-rotation-of-kic-9267654}{%
\section{Case Study: Envelope Rotation of KIC 9267654}\label{case-study-envelope-rotation-of-kic-9267654}}

To simplify the rotational inverse problem, red giants in particular are commonly treated in a two-zone model of differential rotation \citep[e.g.][]{klion_diagnostic_2017}, with the core and envelope assumed to be rotating as more or less solid bodies (if at all). In general, fast rotation also induces nonlinear avoided crossings \citep[e.g.][]{ouazzani_rotation_2013, deheuvels_near_2017} when the off-diagonal matrix elements of \cref{eq:bilinear} are large. However, these change the rotational splitting widths only to third order in perturbation theory \citep[e.g.][]{ong_rotation_2022}, so we will, for the sake of argument, assume that the observed rotational splittings are small enough to remain well-described by the linear expression, \cref{eq:perturb}. In that case, the rotational mutiplet width for a mode of mixed gravitoacoustic character in a red giant takes the form \citep{goupil_seismic_2013}
\begin{equation}
    \delta\omega_\text{rot} \sim \zeta\beta_g \left<\Omega_\text{core}\right> + (1-\zeta)\beta_p \left<\Omega_\text{env}\right>.\label{eq:rot}
\end{equation}
Here the angle brackets denote spatial averages over the g- and p-mode cavities, corresponding roughly to the core and envelope --- we leave the nature of this averaging underspecified --- and \(\zeta\) is a mixing fraction: \(\zeta \to 0\) for a pure p-mode, and \(\to 1\) for a pure g-mode. In the asymptotic limit of high radial order, the normalisation constants take on limiting values of \(\beta_p \to 1, \beta_g \to 1 - 1/\ell(\ell+1)\).

As an illustrative example, we examine a model of KIC 9267654: a rapidly rotating first-ascent red giant exhibiting such gravitoacoustic mixed modes at \(\ell = 1\), whose spectroscopic rotational broadening is greater than a putative envelope-averaged rotation rate implied by a combined seismic constraint from only \(\ell = 2\) and \(\ell = 3\) modes, suggesting rotational shear in the envelope \citep{tayar_spinning_2022}. This is a prima facie surprising result, as single-star evolutionary modelling instead favours close to solid-body envelope rotation. Given that these higher-degree modes are of significantly lower disc-integrated visibility than dipole modes, and penetrate less deeply into its interior, efforts are ongoing to measure envelope rotation rates from its dipole modes as well, to more strongly characterise this unusual rotational configuration.

Some of these ongoing studies of KIC 9267654 use the recently developed eMOLA technique \citep{ahlborn_improved_2022}, which is in the OLA family of methods. Because of its simplicity and demonstrated robustness to noise in model tests, we examine it in detail here. However, our results concern the shapes of the kernels themselves, rather than the coefficients used to sum them, and so also impact all other inversion techniques (both OLA and RLS). Unlike standard OLA, eMOLA is intended only to isolate envelope rotation, rather than to localise rotation to a position \(r_0\), and derives inversion coefficients \(\left\{c_i\right\}\) by optimising a penalty function constructed out of the cumulative sensitivity kernels \(\beta_i(r) = \int_0^r \beta_i K(r') \mathrm d r'\). In addition to the usual contributions to the penalty function described above, the \(\left\{c_i\right\}\) are additionally regularised so as to suppress \(\sum_i c_i \beta_i(r_c)\) at some radius \(r_c\) situated just outside of the small radiative core. Such regularisation suppresses sensitivity of the summed kernel to the radiative core, owing to the highly oscillatory nature of the inversion kernels within the g-mode cavity rapidly cancelling each other out when summed and integrated. By comparison with \cref{eq:perturb,eq:rot}, this is equivalent to demanding that \(\left|\sum_i c_i \beta_{g,i} \zeta_i\right|\) be minimised. Mixed dipole modes in red giants specifically are of very high g-mode radial order, so the g-mode normalisation constants \(\beta_{g,i} \sim 1/2\) do not vary significantly from mode to mode, and may be brought out of this sum as an overall multiplicative factor.

\begin{figure*}[htbp]
    \centering
    \annotate{\includegraphics[width=.45\textwidth, trim=0cm .15cm 0cm .25cm, clip]{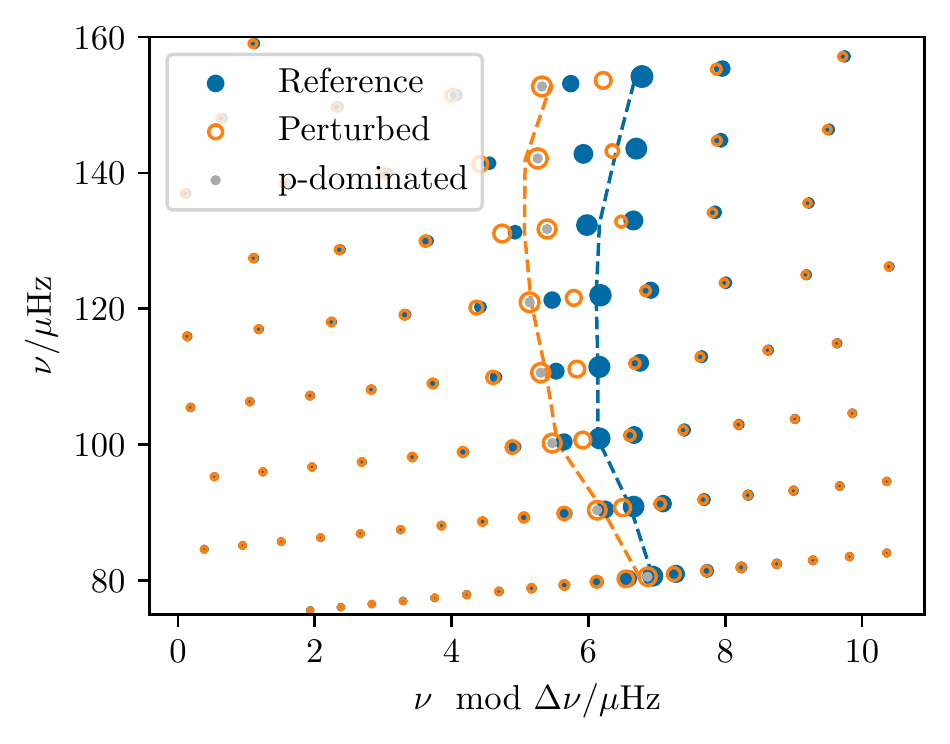}}{\node at (.9,.9){\textbf{(a)}};}
    \annotate{\includegraphics[width=.45\textwidth, trim=0cm .15cm 0cm .25cm, clip]{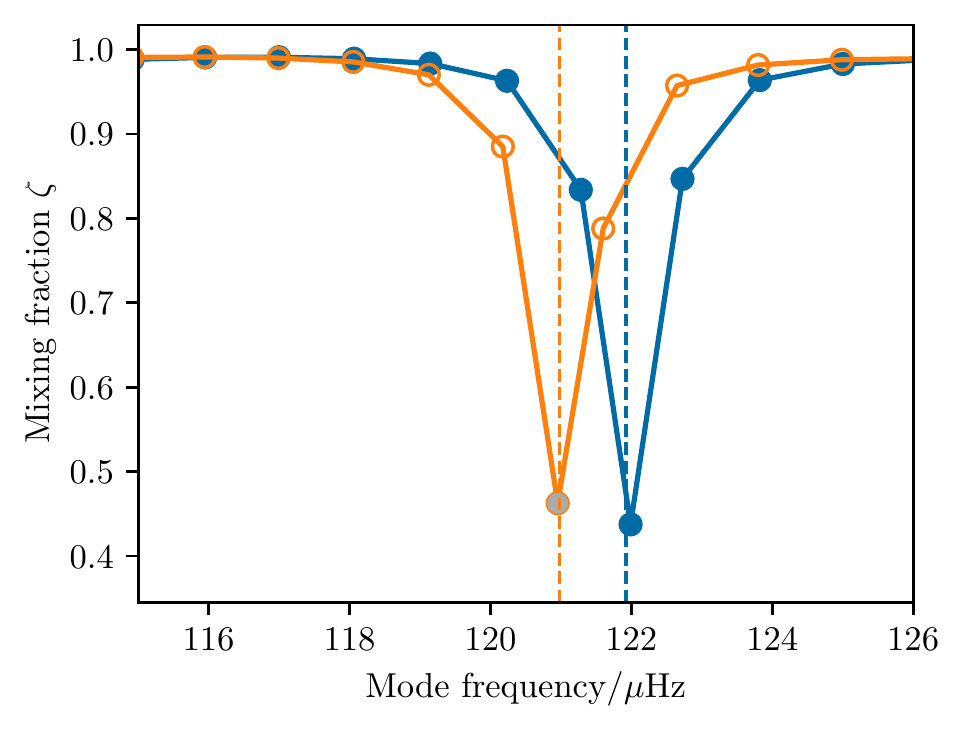}}{\node at (.9,.24){\textbf{(b)}};}
    \annotate{\includegraphics[width=.45\textwidth, trim=0cm .15cm 0cm .25cm, clip]{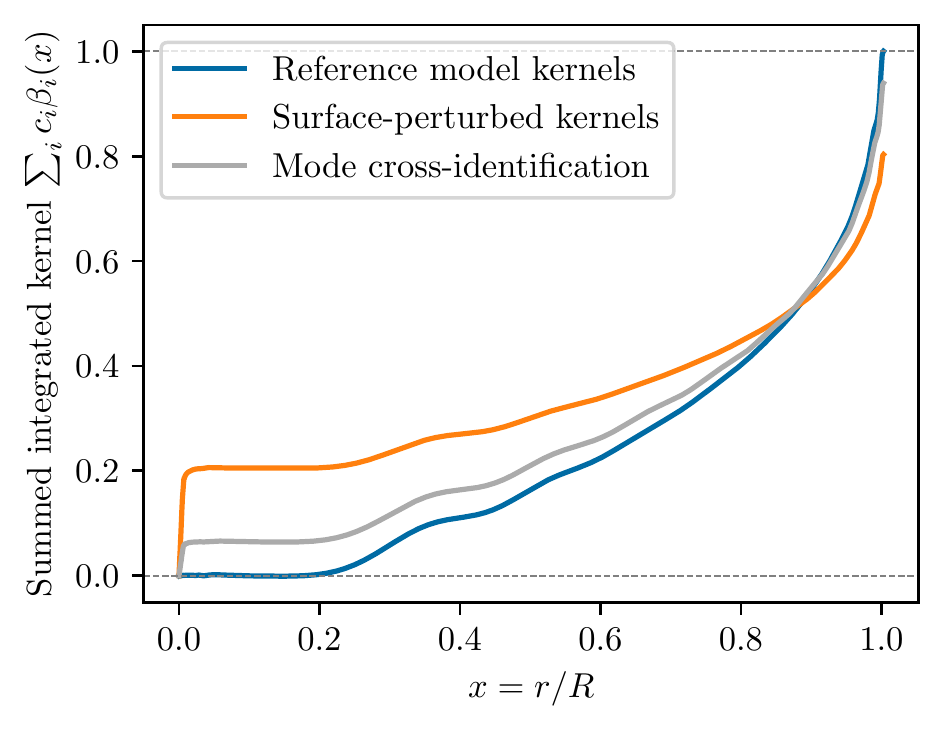}}{\node at (.9,.24){\textbf{(c)}};}
    \annotate{\includegraphics[width=.45\textwidth, trim=0cm .15cm 0cm .25cm, clip]{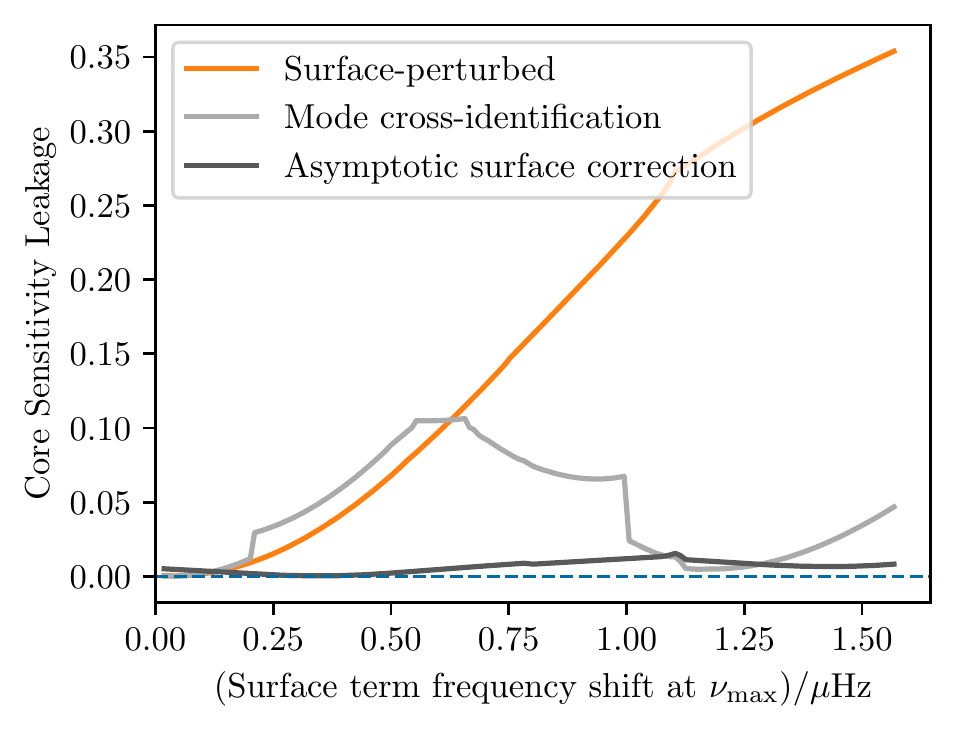}}{\node at (.9,.5){\textbf{(d)}};}
    \caption{In the presence of the surface term, a summed kernel constructed to suppress sensitivity to the core in the reference model will yield nontrivial sensitivity to the core with respect to a surface-perturbed set of modes. \textbf{(a)}: Echelle diagram showing mixed-mode frequencies of a reference model (blue filled circles) of KIC 9267654, and one with a structural perturbation applied to its surface (orange open circles). Markers on the diagram are sized proportionally (by area) to the values of $(1-\zeta)$ for the corresponding modes. Dashed lines indicate the locations of the underlying pure p-modes of the two models. The most p-dominated mixed modes of the surface-perturbed model are indicated with the filled gray circles. \textbf{(b)}: Mixing fractions $\zeta$ for modes in the vicinity of the pure p-mode at $\sim 120\ \mu$Hz. Coloured circles have the same meaning as in panel (a), and are joined with solid lines of matching colour. The most p-dominated surface-perturbed mixed mode is again indicated in gray. Dashed vertical lines denote the locations of the underlying pure p-modes of the models corresponding to each colour. \textbf{(c)}: Summed kernels constructed with the same coefficients, using rotational kernels for the reference (solid blue curve) and surface-perturbed (solid orange curve) models. The latter can be seen to retain sensitivity to the core. Repeating this exercise by cross-identifying modes between models with the lowest values of $\zeta$ can be seen to mitigate this only incompletely. \textbf{(d)}: Residual senstivity to the core exhibited by the summed kernels of the surface-perturbed model, using coefficients of the reference model with nominal mode identification (orange curve), and under cross-identification in the neighbourhood of the most p-dominated modes (light gray curve). Without applying corrections, both can be seen to produce significant leakage. These can be seen to be suppressed by the use of coefficients obtained from the mitigation procedure where asymptotic mode-coupling calculations are performed after applying the surface correction to the underlying pure p-modes of the reference model (dark gray curve). The blue horizontal dashed line shows a notional value of 0. \textbf{(Animation)} The online animated version of this figure shows how the orange and gray objects of panels (a), (b), and (c) respond, as the stellar structure is linearly interpolated from the reference to the surface-perturbed stellar model \edit1{over the course of 7.5 seconds. A vertical gray dotted line indicates the corresponding size of the surface perturbation in panel (d) during this process}.}
    \label{fig:emola}
\end{figure*}

Since all the required properties of the inversion kernels used by the eMOLA procedure are effectively encapsulated in the mixing fraction \(\zeta\), we may recast the condition for optimality as minimizing the expected observational error of the inferred surface rotation rate, subject to constraints both that the summed kernel yields a normalised sensitivity of unity, and that the sum of mixing fractions is 0. In the language of Lagrange multipliers, we optimize an auxiliary penalty function
\begin{equation}
\Lambda = \sum_i c_i^2 \sigma_i^2 - a\left(\sum_i c_i \beta_i - 1\right) - b\left(\sum_i c_i \zeta_i\right),
\end{equation}
where \(\sigma_i\) are the observational errors on the rotational splittings. This gives a closed-form expression for the inversion coefficients \(c_i\) as
\begin{equation}
   c_i = {a \beta_i + b \zeta_i \over 2 \sigma_i^2},\label{eq:emola1}
\end{equation}
where \(a\) and \(b\) solve a \(2\times2\) linear system of equations:
\begin{equation}
    \begin{bmatrix}
    a\\b
    \end{bmatrix}
    =
    \begin{bmatrix}
    \sum_i {\beta_i^2 \over 2 \sigma_i^2} & \sum_i {\beta_i\zeta_i \over 2 \sigma_i^2}\\
    \sum_i {\beta_i\zeta_i \over 2 \sigma_i^2} & \sum_i {\zeta_i^2 \over 2 \sigma_i^2}
    \end{bmatrix}^{-1}
    \begin{bmatrix}
    1\\0
    \end{bmatrix}.\label{eq:emola2}
\end{equation}

We now apply these expressions to an illustrative \mesa~stellar model of KIC 9267654, which was the nearest simultaneous match to its effective temperature, metallicity, Gaia luminosity, radial and quadrupole p-mode frequencies \citep[under the two-term surface-term correction of][]{ball_surface_2014}, and dipole g-mode period spacing \(\Delta\Pi_1\), from the model grid of Lindsay et al.~(submitted to ApJ). Its dipole-mode frequencies and mixing fractions near \(\numax\), computed using the pulsation code \gyre~\citep{townsend_gyre_2013}, are shown in blue in panels (a) and (b) of \cref{fig:emola}. We apply \cref{eq:emola1,eq:emola2} with values of \(\sigma_i\) inversely proportional to a Gaussian envelope function situated at \(\numax\), with width specified by the scaling relation of \citet{mosser_power_2012}. This yields the linear combination of its rotational kernels shown with the blue curve in panel (c) of \cref{fig:emola}, which is insensitive to the core, as desired.

Let us now consider the corresponding summed kernel when the surface structure is slightly different, giving frequency differences from the reference model which are notionally those described by a surface-term correction. Since the true structure and rotational kernels of KIC 9267654 are not available for our inspection, we instead consider a stellar structure that is identical to that of our reference model, except for a small, artificially-induced perturbation to its near-surface layers. As in \obr, we perturb the near-surface layers of the \mesa~model \edit1{(bringing them out of hydrostatic equilibrium)} as \(P(r) \mapsto P(r) \times (1 + \Delta(r))\), with \(\Delta(r) = A \exp\left[-(r/R - 1)^2/2s^2\right]\), and \(s = 0.002\). Using a value of \(A\sim 1\) alters its pure p-modes by about 1 \(\mu\)Hz at \(\numax\), which is roughly in the range of values that \citet{li_surface_2023} find to be needed for similar first-ascent red giants (cf.~their fig.~3). This is also roughly the size of the correction needed to bring the radial and quadrupole modes of this model into agreement with their observed values. We emphasise that in all other respects the inner structure of the perturbed model is identical to that of the reference; in particular the interior near-core layers, which determine the strength of coupling between the p- and g-modes, are untouched.

The resulting mode frequencies and mixing fractions \edit1{(recomputed with \gyre\ using identical settings as for the reference model)} are shown in orange in panels (a) and (b) of \cref{fig:emola}. Importantly, as described in \obr, the mixing fractions can be seen to change under the action of the surface term. This nonlinear interaction between the structure and mode frequencies also significantly changes the shapes of the inversion kernels compared to the reference model. Thus, if a linear combination of kernels of the reference model suppresses sensitivity to the core, that same linear combination of kernels from the surface-perturbed structure will fail to do so. This sum of the surface-perturbed kernels is depicted with the solid orange curve in panel (c) of \cref{fig:emola}. The corresponding linear combination of observed rotational splittings is thus nontrivially sensitive to core rotation, and ought not be interpreted as signifying only an isolated envelope rotation rate. In the depicted configuration, \cref{fig:emola}c shows that this adds up to 20\% of the core rotation rate (on top of overall reduced sensitivity to the envelope of only 60\%); we will refer to this as a 20\% ``core sensitivity leakage'' in our following discussion. \edit1{Given coefficients $\{c_{i, \text{ref}}\}$ of the reference model and mixing fractions $\zeta_{i, \text{targ}}$ and sensitivity factors $\beta_{i, \text{targ}}$ of the target model, we compute this leakage $L$ as
\begin{equation}
    L \sim \sum_i c_{i, \text{ref}}\beta_{i, \text{targ}} \zeta_{i, \text{targ}}.
\end{equation}
}

A further complication to this situation is that, in actual observations, the identities of the modes (in particular their radial orders \(n_{pg}\)) are not known in advance. While we have assigned inversion coefficients from the reference model perfectly to modes with the same radial orders in the perturbed model (as is required by \cref{eq:perturb}), \cref{fig:emola}b shows that, under the action of a sufficiently large surface perturbation, the most p-dominated mixed mode in the reference model may not have the same radial order as that of the observed star. This being the case, one might argue that cross-identifying modes in assigning coefficients --- so that the most p-dominated (and thus highest-amplitude) observed modes and rotational splittings are assigned the inversion coefficients associated with the most p-dominated modes in the reference model --- might at least mitigate the issue. For the particular configuration shown, this is indeed the case. For illustration, we derive coefficients from the reference model for only the most p-dominated mixed modes, and the 2 nearest ones on each side. We then assign those coefficients to the most p-dominated mixed modes of the surface-perturbed model (indicated in gray in \cref{fig:emola}a and b) and likewise adjacent modes. The core sensitivity leakage of the resulting summed surface-perturbed kernel (gray curve of \cref{fig:emola}c) can be seen to be reduced, although not entirely eliminated, for this particular perturbed structure.

However, this introduces further problems, in the form of significant discontinuities and nonmonotonicity. We illustrate this with the orange and gray curves in \cref{fig:emola}d, which show how the core sensitivity leakage responds in both cases, as we smoothly interpolate the target model from the reference to the surface-perturbed structure. As the two structures diverge, the core sensitivity leakage under perfect mode identification increases in a monotonic fashion (orange curve), but that from cross-identifying modes does not, instead suffering unpredictable discontinuities with each discrete crossover of mode identification. The mitigation this affords is also in some cases actually counterproductive, leaking substantially more than direct mode identification. These are clearly undesirable properties of any mitigation procedure, and so such cross-identification is not a viable path forward.

\hypertarget{mitigation-strategy}{%
\section{Mitigation Strategy}\label{mitigation-strategy}}

\begin{figure}
\centering
\includegraphics{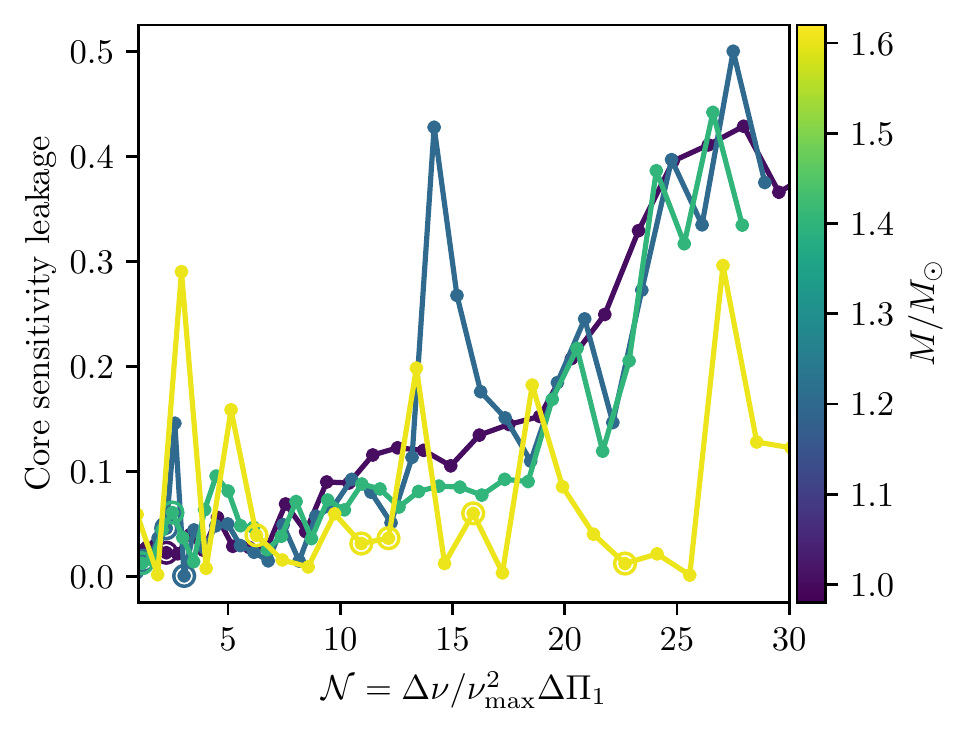}
\caption{eMOLA core sensitivity leakages computed for several tracks of solar-composition \mesa~models, assuming perfect mode identification, and assuming the surface term to take the parameterisation of \citet{ball_surface_2014}, with coefficients varying over evolution in the manner prescribed by \citet{li_surface_2023}. These are plotted as functions of the relative g-mode density \(\mathcal{N} = \Dnu/\numax^2\Delta\Pi_1\): a proxy for evolution up the red giant branch. Models of each evolutionary track, indicated using filled circles coloured by mass, are shown connected with lines. Vertical positions of the markers indicate only the absolute value of the core leakage, with negative values indicated with additional open circles around individual markers. \label{fig:leakage-evol}}
\end{figure}

So far, we have been restricted to qualitative analysis of an illustrative stellar model of a single star. However, this qualitative discussion indicates the existence of measurement systematics for envelope rotation, which we will now quantify. We consider the same set of evolutionary models as the hare-and-hounds exercise of \citet{ong_rotation_2023}, consisting of several evolutionary tracks of \mesa~models at different masses, solar composition, and solar-calibrated mixing length. Their p- and g-mode frequencies have been calculated using the prescription of \citet{ong_semianalytic_2020}. For each of these models, we compute the core leakages that would result from assigning the eMOLA coefficients of \cref{eq:emola1}, as computed from the nominal mixed modes, to the modes and mixing fractions of the surface-corrected mixed modes, corrected according to the prescription of \obr, and with coefficients for the pure p-modes of the two-term \citet{ball_surface_2014} parameterisation computed from the calibration of \citet{li_surface_2023}. These are shown as a function of the relative g-mode density \(\mathcal{N} = \Dnu/\numax^2\Delta\Pi_1\), a proxy for evolution up the red giant branch, in \cref{fig:leakage-evol}. We restrict our attention to \(1 < \mathcal{N} < 30\), as the observed oscillations are primarily p-dominated modes at lower \(\mathcal{N}\), and mode mixing ceases to be an observational concern for more evolved stars at higher \(\mathcal{N}\). While the amount of leakage appears to fluctuate significantly over the course of evolution (as the precise amount of leakage will depend on the relative positions of the underlying p- and g-modes), potentially even becoming negative, its severity can nonetheless be seen roughly to increase with evolution. This is in line with theoretical considerations, given that \citet{li_surface_2023} finds the size of the surface term to be a roughly constant fraction of \(\Dnu\), while the ratio of the nonlinearity threshhold of \obr~to \Dnu~decreases with increasing \(\mathcal{N}\).

The cores of red giants are known to rotate more rapidly, by an order of magnitude or more, than their envelopes. Even a small, 10\%, leakage of core sensitivity --- well within the range of plausible values shown here --- would then imply an amount of contamination comparable to the entire existing reported dipole-mode envelope rotation rates for red giants like KIC 9267654 \citep{triana_internal_2017}, assuming these were not themselves contaminated in this fashion. Conversely, it is plausible that existing reported dipole-mode inferences of envelope rotation in such red giants --- or even of counterrotation \citep[e.g.][]{deheuvels_seismic_2012, deheuvels_seismic_2014}, given the appearance of negative values for the core leakage in our modelling exercise --- may be dominated by such contamination to begin with. Thus, the surface term is a hitherto unaccounted-for, but potentially dominant, source of systematic observational error in these rotational inferences. Clearly, there is a pressing need for mitigating measures.

\begin{figure}
\centering
\includegraphics{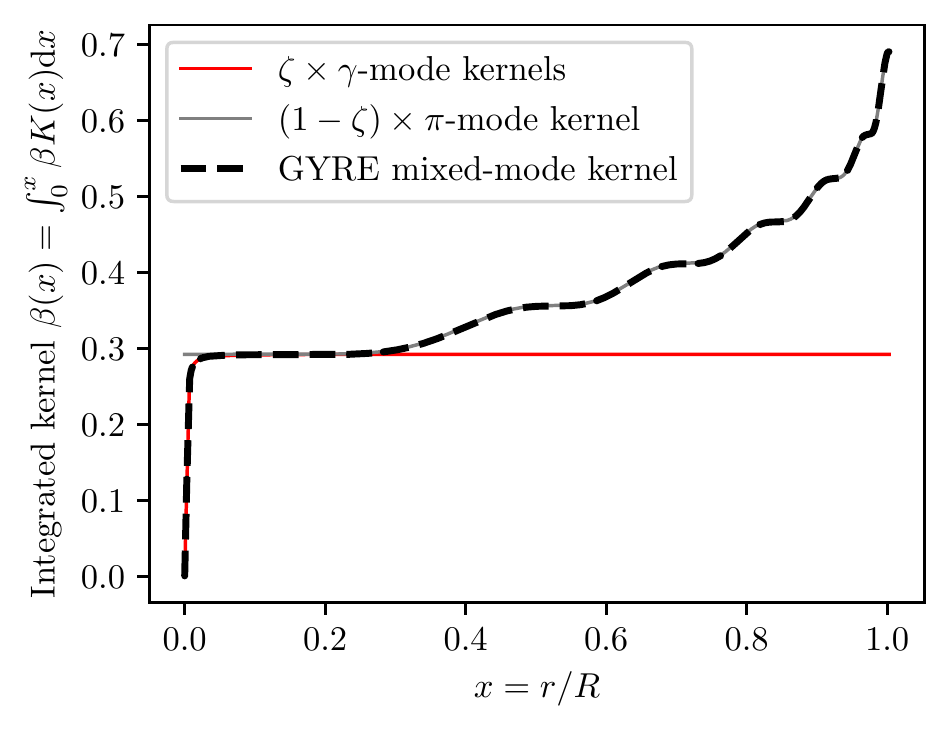}
\caption{Mixed-mode rotational kernels may be constructed as linear combinations of pure p- and g-mode rotational kernels. Here, the cumulative integral for a mixed-mode rotational kernel of the surface-perturbed model (black dashed line), computed directly from its eigenfunction returned from \gyre, is compared against a combination of that for an averaged pure g-mode rotational kernel (red solid line) scaled by the mixing fraction, with that for the nearest pure p-mode rotational kernel (gray solid line), scaled complementarily, returned from the reference model.\label{fig:recombination}}
\end{figure}

We now outline a path forward. \obr~described how standard surface-term corrections, applied to the pure p-modes of a stellar model \citep[e.g.~as computed through the \(\pi/\gamma\) decomposition scheme of of][]{ong_semianalytic_2020}, yield surface-corrected mixed modes through mode-coupling calculations in which the mixed-mode eigenfunctions \(\vec{\xi}\) are expressed as linear combinations of pure p- and g-mode eigenfunctions. For mixed modes near each pure p-mode, only that p-mode contributes appreciably to this combination: to a good approximation,
\begin{equation}
    \vec{\xi} \sim a_p \vec{\xi}_p + \sum_j a_{g, j} \vec{\xi}_{g, j}.\label{eq:combination}
\end{equation}
The coefficients of these linear combinations yield mixing fractions, as under unit normalisation (\(\left<\vec{\xi}, \vec{\xi}\right> = 1\)), they satisfy \(\sum_j |a_{g, j}|^2 \sim \zeta\) and \(|a_p|^2 \sim 1 - \zeta\). Given that the rotational kernels are produced from a bilinear form (of the kind specified in \cref{eq:bilinear}) acting on the eigenfunctions as \(\beta K = R[\vec{\xi}, \vec{\xi}]\), we then have
\begin{equation}
\begin{aligned}
\beta K = R[\vec{\xi},\vec{\xi}] &= R\left[\sum_j a_{g,j} \vec{\xi}_{g,j} + a_p \vec{\xi}_p,\sum_j a_{g,j} \vec{\xi}_{g,j} + a_p \vec{\xi}_p\right]\\
&\sim a_p^2 R[\vec{\xi}_p, \vec{\xi}_p] + \sum_j a_{g,j}^2 K[\vec{\xi}_{g,j}, \vec{\xi}_{g,j}]\\
&\sim (1-\zeta) \beta_p K_p + \zeta \left<\beta_g K_g\right>.\label{eq:recombination}
\end{aligned}  
\end{equation}
Here we neglect cross-terms of the form \(R[\vec{\xi}_p, \vec{\xi}_g]\), as \citet{ong_rotation_2022} show that doing so, in this transformed basis of pure p- and g-modes, remains a good approximation even in the presence of nonlinear rotational avoided crossings in the natural basis of mixed modes. Angle brackets here represent a heuristic weighted average with respect to the squared mixing coefficients \(|a_{g,j}|^2\). In words: we may well approximate the mixed-mode rotational kernels as linear combinations of pure p- and g-mode rotational kernels. Details for their computation, given a set of pure p- and g-mode eigenfunctions, are provided in \citet{ong_rotation_2022}. We illustrate this construction in \cref{fig:recombination}.

With this in hand, we may now propose our mitigation procedure. Given a reference model, one:

\begin{enumerate}
\def\labelenumi{\arabic{enumi}.}
\tightlist
\item
  Computes the pure p- and g-mode frequencies and eigenfunctions from the reference model \citep[e.g.~using the prescription of][]{ong_semianalytic_2020},
\item
  Applies a surface term correction to the pure p-modes,
\item
  Performs mode-coupling calculations to obtain mixed-mode frequencies and mixing fractions, and
\item
  Constructs mixed-mode rotational kernels of the form specified by \cref{eq:recombination}. Subsequent inversion techniques are then to operate with respect to these reconstituted kernels.
\end{enumerate}

Under the prescription of \obr, step 3 requires solving for the matrix eigenvectors of a generalised Hermitian eigenvalue problem. However, the large rank of these matrices, particularly for evolved red giants, may render this procedure prohibitively expensive. As an alternative, one may instead solve for the roots of the asymptotic characteristic function
\begin{equation}
    F(\nu) = \tan \Theta_p(\nu) \tan \Theta_g(\nu) - q(\nu), \text{ such that } F(\nu_\text{mixed}) = 0, \label{eq:eig}
\end{equation}
which arises from Jeffreys-Wentzel--Kramers--Brillouin (JWKB) analysis of the mixed-mode problem \citep{shibahashi_modal_1979, unno_nonradial_1989}. \citet{ong_rotation_2023} derive a map from the characteristic function of the matrix problem to this JWKB characteristic function, as well as expressions for \(\Theta_p\), \(\Theta_g\), and \(q\), in terms of the nonasymptotic pure p- and g-mode eigensystems. Given roots \(\nu_i\) of \cref{eq:eig}, the mixing fractions may then be found as the values of an asymptotic mixing function: \(\zeta_i = \zeta_p(\nu_i)\). The averaging over g-mode mixing coefficients \(a_{g,j}^2\) of \cref{eq:recombination} may also be replaced with an average over Brillouin-Wigner resonance factors, as \(a_{g,j}(\nu_i) \propto 1/\left(\nu_{g,j}^2 - \nu_i^2\right)\) --- we derive this explicitly in \autoref{sec:bw}. These approximate closed-form expressions, which hold best in the first-ascent-giant limit of low \(q\), are much faster than the matrix calculation.

We show with the dark gray curve in \cref{fig:emola}d the core sensitivity leakage from summing kernels of the surface-perturbed model, using eMOLA coefficients obtained after applying our mitigation to the reference model, and using the above approximate JWKB expressions rather than solving the matrix eigenvalue problem. This simplified procedure can still be seen to efficaciously suppress such leakage without necessitating more expensive matrix calculations. We propose its adoption as a fast and easily implementable preprocessing step to the kernels used in existing code for rotational inversions with mixed modes.

\hypertarget{discussion-and-conclusion}{%
\section{Discussion and Conclusion}\label{discussion-and-conclusion}}

Corrections for the surface term are not ordinarily considered necessary for rotational inversions. Even where such corrections have been applied, they have been thought to affect only the mode frequencies, and so all kernels that have been used for existing inversions remain those of the uncorrected modes in the reference models. We have now demonstrated how the action of the surface term changes the mixing fraction, and thus the shape of the rotational kernels, for mixed gravitoacoustic modes. As a result, this may have caused previous estimates of red-giant envelope rotation rates from mixed-mode asteroseismic inversions to have been unintentionally contaminated by core rotation. The potential scope of this issue is very broad, as it impacts effectively every existing mixed-mode rotational inversion. These unaccounted surface effects may explain the widely varying observational discrepancies --- of comparable size to the reported values --- between envelope rotation rates estimated from inversion techniques, compared to diagnostics derived from only the asymptotic theory without reference to stellar models \citep[e.g.][]{goupil_seismic_2013, klion_diagnostic_2017, triana_internal_2017, gehan_core_2018, mosser_period_2018}. While our above qualitative discussion was restricted to demonstrating core-rotation leakage into estimates of envelope rotation, it also conversely implies the potential existence of envelope-rotation leakage into existing estimates of core (differential) rotation.

We have described an explicit procedure to mitigate this issue. One first applies surface-term corrections to the mixed modes, via the generalisation proposed in \obr, to obtain both surface-corrected mode frequencies, as well as surface-corrected mixing coefficients between the pure p- and g-modes. We have furthermore derived expressions for these mixing coefficients in terms of the commonly-used approximate JWKB formalism of mixed-mode coupling, to accelerate the matrix calculations otherwise required by the prescription of \obr. One then uses these mixing coefficients to combine pure p- and g-mode rotational kernels, of the kind described in \citet{ong_rotation_2022}, back into surface-corrected mixed-mode rotational kernels.

Similar considerations also apply to mixed-mode inversions for stellar structure, and may contribute to the so far inconclusive nature of attempts at this for stars more evolved than subgiants \citep[and pers. comm.]{bellinger_asteroseismic_2021}. However, the procedure that we have proposed here may not be directly applicable to the structure problem. Here, the shapes of the rotational kernels are changed only in response to perturbations to the stellar structure near the surface, which changes we reverse upon applying a nonlinear surface correction. Should the reference and target model differ in the stellar interior, and specificially in the evanescent region between the p- and g-mode cavities, then we may not necessarily be able to drop the p-g cross-terms that would appear in the structural equivalent to \cref{eq:recombination}. Equivalently, a perturbation to these regions may change the coupling strength between the two mode cavities, causing even the surface-corrected mixing fractions of the reference model to differ from that of the observed star. Better understanding the limitations of this procedure will be a crucial first step towards generalising it, and making further progress on the mixed-mode structure inversion problem. Corollarily, stellar models used for even the rotational inverse problem have to be constructed so that their gravitoacoustic coupling strengths match those of the observed stars.

While asteroseismic inversions are generally quite involved, our proposed mitigation requires only an additional preprocessing step to be applied to the inversion kernels --- that is, only the penultimate step of a long analysis needs to be modified. This being the case, any revisions to these studies should be quick and unburdensome, so long as the original stellar models used for inversion remain available for reanalysis. To assist in these efforts, we have made our Python implementation of this mitigating prescription publicly accessible. This will in turn have immediate and astrophysically meaningful consequences. Seismic rotation measurements are how red giants inform the angular-momentum transport problem \citep[e.g.][]{aerts_angular_2019, fuller_slowing_2019}, and so such revisions may substantially modify our current understanding of it. We await these new findings with great excitement and anticipation.

% \begin{acknowledgements}

We thank D. Huber, J. van Saders, and S. Basu, for constructive feedback and helpful discussions, and F. Ahlborn for insightful and enjoyable discussions about the eMOLA technique. JMJO acknowledges support from NASA through the NASA Hubble Fellowship grant HST-HF2-51517.001, awarded by STScI, which is operated by the Association of Universities for Research in Astronomy, Incorporated, under NASA contract NAS5-26555. Our Python implementation of the procedure described here, along with our input namelist files and the depicted reference \mesa\ model, is freely available on Zenodo at \dataset[doi:10.5281/zenodo.8433219]{\doi{10.5281/zenodo.8433219}}.

\software{NumPy \citep{numpy}, SciPy stack \citep{scipy}, AstroPy \citep{astropy:2013,astropy:2018,astropy:2022}, Pandas \citep{pandas}, \mesa\ \citep{mesa_paper_1,mesa_paper_2,mesa_paper_3,mesa_paper_4,mesa_paper_5}, \gyre\ \citep{townsend_gyre_2013}.}
% \end{acknowledgements}

\appendix
\setcounter{table}{0}
\renewcommand{\thetable}{A\arabic{table}}

\setcounter{figure}{0}
\renewcommand{\thefigure}{A\arabic{figure}}

\hypertarget{brillouin-wigner-perturbation-theory}{%
\section{\texorpdfstring{Brillouin-Wigner perturbation theory\label{sec:bw}}{Brillouin-Wigner perturbation theory}}\label{brillouin-wigner-perturbation-theory}}

Expressions like \cref{eq:expand} emerge from standard Rayleigh-Schrödinger perturbation theory \citep[e.g.][]{schrodinger_quantisierung_1926}, which is used when both the perturbed eigenvalues and eigenvectors of a linear system are to be found simultaneously from the unperturbed eigensystem. However, when the perturbed eigenvalues are already known, one may instead make use of Brillouin-Wigner perturbation theory, which then yields expressions for the perturbed eigenvectors that are both more accurate (as the resulting perturbative expansion converges more rapidly), and also are far simpler, especially at higher order. We refer the reader to e.g.~§3.5 of \citet{killingbeck_perturbation_1977} for a more thorough review regarding the properties of this approach to perturbation theory. However, since this is (to our knowledge) its first application in an asteroseismic context, we will also provide a rough overview of its main points here for quick reference.

Consider the eigenvectors \(\ket{n}\) of a linear operator \(\hat{H} = \hat{H}_0 + \lambda \hat{V}\) with small \(\lambda\). We assume that we know in advance both the eigenvalues of \(\hat{H_0}\), which are \(\epsilon_n\), and of \(\hat{H}\), which are \(E_n\). We wish to expand \(\ket{n}\) in powers of \(\lambda\) as \(\ket{n} \sim \ket{n^{(0)}} + \lambda \ket{n^{(1)}} + \lambda^2 \ket{n^{(2)}} + \ldots\), in the usual fashion. For convenience, we will operate under the convention that \(\braket{n^{(0)}|n} = 1\), so that the unperturbed eigenvector \(\ket{n^{(0)}}\) only enters the expansion to zeroth order. By definition,
\begin{equation}
\begin{aligned}
    E_n\braket{m^{(0)}|n} &= \braket{m^{(0)}|\hat{H}|n} \\&= \braket{m^{(0)}|\left(\hat{H}_0 + \lambda \hat{V}\right)|n} \\&= \epsilon_m\braket{m^{(0)}|n} + \lambda \braket{m^{(0)}|\hat{V}|n}
    \\\implies E_n \sum_k \lambda^k\braket{m^{(0)}|n^{(k)}} &= \sum_k \lambda^k \left(\epsilon_m\braket{m^{(0)}|n^{(k)}} + \braket{m^{(0)}|\hat{V}|n^{(k-1)}}\right).
\end{aligned}
\end{equation}
This gives us a recurrence relation for \(k > 0\), \(m \ne n\) that \(\braket{m^{(0)}|n^{(k)}} = \braket{m^{(0)}|\hat{V}|n^{(k-1)}} / (E_n - \epsilon_m)\), whence it follows immediately that
\begin{equation}
\begin{aligned}
    \ket{n} &\sim \ket{n^{(0)}} + \lambda \sum_{m\ne n}{\braket{m^{(0)}|\hat{V}|n^{(0)}} \over E_n-\epsilon_m} \ket{m^{(0)}} \\&+ \lambda^2 \sum_{m_1\ne n}\sum_{m_2\ne n}{\braket{m_2^{(0)}|\hat{V}|m_1^{(0)}}\braket{m_1^{(0)}|\hat{V}|n^{(0)}} \over (E_n-\epsilon_{m_2})(E_n-\epsilon_{m_1})}\ket{m_2^{(0)}} + \ldots
\end{aligned}\label{eq:bw}
\end{equation}
These expressions are understood to either be additionally scaled by an overall factor, or else to contain additional contributions parallel to \(\ket{n^{(0)}}\) at each succeeding order, if one wished to instead give unit normalisation for \(\ket{n}\).

In our application of Brillouin-Wigner perturbation theory, the perturbed frequency eigenvalues are the mixed-mode frequencies supplied from solving for the roots of the asymptotic characteristic function of \cref{eq:eig}. \citet{ong_rotation_2023} demonstrate that these correspond to the eigenvalues of the matrix eigenvalue problem, in the case of weak coupling between the pure p- and g-modes. Thus, for our purposes, our ``unperturbed'' system consists of the pure p- and g-modes, and the perturbation to the system, represented by \(\hat{V}\) in \cref{eq:bw}, is the coupling between them that produces mixed modes. This coupling has the further property of only being nonzero between p- and g-modes; that is, there is none between pure p-modes and other pure p-modes, or between pure g-modes and other pure g-modes. Finally, we may assume that the coupling strength may be treated as a smooth function of frequency. Thus, \cref{eq:bw} gives to leading order, and under unit normalisation, that for a g-dominated mixed mode,
\begin{equation}
    \vec{\xi}_{i} \sim \sqrt{\zeta_i}\ \vec{\xi}_{g, i} + \sqrt{1 - \zeta_i} \sum_j {A_i \over \omega_{p,j}^2 - \omega_{i}^2} \vec{\xi}_{p,j},
\end{equation}
where the \(A_i\) are normalisation factors such that \(\sum_j \left|A_i \left/ \left(\omega_{p,j}^2 - \omega_{i}^2\right)\right.\right|^2 = 1\). Conversely, for a p-dominated mixed mode,
\begin{equation}
    \vec{\xi}_{i} \sim \sqrt{1 - \zeta_i}\ \vec{\xi}_{p, i} + \sqrt{\zeta_i} \sum_j {B_i \over \omega_{g,j}^2 - \omega_{i}^2} \vec{\xi}_{g,j},
\end{equation}
subject to the same kind of normalisation for the pure-g-mode component. It is this latter first-order expression which we use in \autoref{mitigation-strategy}.

  \bibliography{biblio.bib}

\begin{thebibliography}{}
\expandafter\ifx\csname natexlab\endcsname\relax\def\natexlab#1{#1}\fi
\providecommand{\url}[1]{\href{#1}{#1}}
\providecommand{\mhref}[2]{\href{#1}{\color{magenta}#2}}
\providecommand{\dodoi}[1]{doi:~\href{http://doi.org/#1}{\nolinkurl{#1}}}
\providecommand{\doeprint}[1]{\href{http://ascl.net/#1}{\nolinkurl{http://ascl.net/#1}}}
\providecommand{\doarXiv}[1]{\href{https://arxiv.org/abs/#1}{\nolinkurl{https://arxiv.org/abs/#1}}}

\bibitem[{{Aerts} {et~al.}(2019){Aerts}, {Mathis}, \&
  {Rogers}}]{aerts_angular_2019}
{Aerts}, C., {Mathis}, S., \& {Rogers}, T.~M. 2019,
  {\mhref{http://doi.org/10.1146/annurev-astro-091918-104359}{\araa}},
  {\href{https://ui.adsabs.harvard.edu/abs/2019ARA&A..57...35A}{57}}{\href{https://ui.adsabs.harvard.edu/abs/2019ARA&A..57...35A}{,
  35}}

\bibitem[{{Ahlborn} {et~al.}(2022){Ahlborn}, {Bellinger}, {Hekker}, {Basu}, \&
  {Mokrytska}}]{ahlborn_improved_2022}
{Ahlborn}, F., {Bellinger}, E.~P., {Hekker}, S., {Basu}, S., \& {Mokrytska}, D.
  2022, {\mhref{http://doi.org/10.1051/0004-6361/202142510}{\aap}},
  {\href{https://ui.adsabs.harvard.edu/abs/2022A&A...668A..98A}{668}}{\href{https://ui.adsabs.harvard.edu/abs/2022A&A...668A..98A}{,
  A98}}

\bibitem[{{Astropy Collaboration} {et~al.}(2013){Astropy Collaboration},
  {Robitaille}, {Tollerud}, {Greenfield}, {Droettboom}, {Bray}, {Aldcroft},
  {Davis}, {Ginsburg}, {Price-Whelan}, {Kerzendorf}, {Conley}, {Crighton},
  {Barbary}, {Muna}, {Ferguson}, {Grollier}, {Parikh}, {Nair}, {Unther},
  {Deil}, {Woillez}, {Conseil}, {Kramer}, {Turner}, {Singer}, {Fox}, {Weaver},
  {Zabalza}, {Edwards}, {Azalee Bostroem}, {Burke}, {Casey}, {Crawford},
  {Dencheva}, {Ely}, {Jenness}, {Labrie}, {Lim}, {Pierfederici}, {Pontzen},
  {Ptak}, {Refsdal}, {Servillat}, \& {Streicher}}]{astropy:2013}
{Astropy Collaboration}, {Robitaille}, T.~P., {Tollerud}, E.~J., {et~al.} 2013,
  {\mhref{http://doi.org/10.1051/0004-6361/201322068}{\aap}},
  {\href{http://adsabs.harvard.edu/abs/2013A%26A...558A..33A}{558}}{\href{http://adsabs.harvard.edu/abs/2013A%26A...558A..33A}{,
  A33}}

\bibitem[{{Astropy Collaboration} {et~al.}(2018){Astropy Collaboration},
  {Price-Whelan}, {Sip{\H{o}}cz}, {G{\"u}nther}, {Lim}, {Crawford}, {Conseil},
  {Shupe}, {Craig}, {Dencheva}, {Ginsburg}, {VanderPlas}, {Bradley},
  {P{\'e}rez-Su{\'a}rez}, {de Val-Borro}, {Aldcroft}, {Cruz}, {Robitaille},
  {Tollerud}, {Ardelean}, {Babej}, {Bach}, {Bachetti}, {Bakanov}, {Bamford},
  {Barentsen}, {Barmby}, {Baumbach}, {Berry}, {Biscani}, {Boquien}, {Bostroem},
  {Bouma}, {Brammer}, {Bray}, {Breytenbach}, {Buddelmeijer}, {Burke},
  {Calderone}, {Cano Rodr{\'\i}guez}, {Cara}, {Cardoso}, {Cheedella}, {Copin},
  {Corrales}, {Crichton}, {D'Avella}, {Deil}, {Depagne}, {Dietrich}, {Donath},
  {Droettboom}, {Earl}, {Erben}, {Fabbro}, {Ferreira}, {Finethy}, {Fox},
  {Garrison}, {Gibbons}, {Goldstein}, {Gommers}, {Greco}, {Greenfield},
  {Groener}, {Grollier}, {Hagen}, {Hirst}, {Homeier}, {Horton}, {Hosseinzadeh},
  {Hu}, {Hunkeler}, {Ivezi{\'c}}, {Jain}, {Jenness}, {Kanarek}, {Kendrew},
  {Kern}, {Kerzendorf}, {Khvalko}, {King}, {Kirkby}, {Kulkarni}, {Kumar},
  {Lee}, {Lenz}, {Littlefair}, {Ma}, {Macleod}, {Mastropietro}, {McCully},
  {Montagnac}, {Morris}, {Mueller}, {Mumford}, {Muna}, {Murphy}, {Nelson},
  {Nguyen}, {Ninan}, {N{\"o}the}, {Ogaz}, {Oh}, {Parejko}, {Parley}, {Pascual},
  {Patil}, {Patil}, {Plunkett}, {Prochaska}, {Rastogi}, {Reddy Janga},
  {Sabater}, {Sakurikar}, {Seifert}, {Sherbert}, {Sherwood-Taylor}, {Shih},
  {Sick}, {Silbiger}, {Singanamalla}, {Singer}, {Sladen}, {Sooley},
  {Sornarajah}, {Streicher}, {Teuben}, {Thomas}, {Tremblay}, {Turner},
  {Terr{\'o}n}, {van Kerkwijk}, {de la Vega}, {Watkins}, {Weaver}, {Whitmore},
  {Woillez}, {Zabalza}, \& {Astropy Contributors}}]{astropy:2018}
{Astropy Collaboration}, {Price-Whelan}, A.~M., {Sip{\H{o}}cz}, B.~M., {et~al.}
  2018, {\mhref{http://doi.org/10.3847/1538-3881/aabc4f}{\aj}},
  {\href{https://ui.adsabs.harvard.edu/abs/2018AJ....156..123A}{156}}{\href{https://ui.adsabs.harvard.edu/abs/2018AJ....156..123A}{,
  123}}

\bibitem[{{Astropy Collaboration} {et~al.}(2022){Astropy Collaboration},
  {Price-Whelan}, {Lim}, {Earl}, {Starkman}, {Bradley}, {Shupe}, {Patil},
  {Corrales}, {Brasseur}, {N{\"o}the}, {Donath}, {Tollerud}, {Morris},
  {Ginsburg}, {Vaher}, {Weaver}, {Tocknell}, {Jamieson}, {van Kerkwijk},
  {Robitaille}, {Merry}, {Bachetti}, {G{\"u}nther}, {Aldcroft},
  {Alvarado-Montes}, {Archibald}, {B{\'o}di}, {Bapat}, {Barentsen},
  {Baz{\'a}n}, {Biswas}, {Boquien}, {Burke}, {Cara}, {Cara}, {Conroy},
  {Conseil}, {Craig}, {Cross}, {Cruz}, {D'Eugenio}, {Dencheva}, {Devillepoix},
  {Dietrich}, {Eigenbrot}, {Erben}, {Ferreira}, {Foreman-Mackey}, {Fox},
  {Freij}, {Garg}, {Geda}, {Glattly}, {Gondhalekar}, {Gordon}, {Grant},
  {Greenfield}, {Groener}, {Guest}, {Gurovich}, {Handberg}, {Hart},
  {Hatfield-Dodds}, {Homeier}, {Hosseinzadeh}, {Jenness}, {Jones}, {Joseph},
  {Kalmbach}, {Karamehmetoglu}, {Ka{\l}uszy{\'n}ski}, {Kelley}, {Kern},
  {Kerzendorf}, {Koch}, {Kulumani}, {Lee}, {Ly}, {Ma}, {MacBride}, {Maljaars},
  {Muna}, {Murphy}, {Norman}, {O'Steen}, {Oman}, {Pacifici}, {Pascual},
  {Pascual-Granado}, {Patil}, {Perren}, {Pickering}, {Rastogi}, {Roulston},
  {Ryan}, {Rykoff}, {Sabater}, {Sakurikar}, {Salgado}, {Sanghi}, {Saunders},
  {Savchenko}, {Schwardt}, {Seifert-Eckert}, {Shih}, {Jain}, {Shukla}, {Sick},
  {Simpson}, {Singanamalla}, {Singer}, {Singhal}, {Sinha}, {Sip{\H{o}}cz},
  {Spitler}, {Stansby}, {Streicher}, {{\v{S}}umak}, {Swinbank}, {Taranu},
  {Tewary}, {Tremblay}, {de Val-Borro}, {Van Kooten}, {Vasovi{\'c}}, {Verma},
  {de Miranda Cardoso}, {Williams}, {Wilson}, {Winkel}, {Wood-Vasey}, {Xue},
  {Yoachim}, {Zhang}, {Zonca}, \& {Astropy Project
  Contributors}}]{astropy:2022}
{Astropy Collaboration}, {Price-Whelan}, A.~M., {Lim}, P.~L., {et~al.} 2022,
  {\mhref{http://doi.org/10.3847/1538-4357/ac7c74}{\apj}},
  {\href{https://ui.adsabs.harvard.edu/abs/2022ApJ...935..167A}{935}}{\href{https://ui.adsabs.harvard.edu/abs/2022ApJ...935..167A}{,
  167}}

\bibitem[{{Backus} \& {Gilbert}(1968)}]{backus_resolving_1968}
{Backus}, G., \& {Gilbert}, F. 1968,
  {\mhref{http://doi.org/10.1111/j.1365-246X.1968.tb00216.x}{Geophysical
  Journal}},
  {\href{https://ui.adsabs.harvard.edu/abs/1968GeoJ...16..169B}{16}}{\href{https://ui.adsabs.harvard.edu/abs/1968GeoJ...16..169B}{,
  169}}

\bibitem[{{Ball} \& {Gizon}(2014)}]{ball_surface_2014}
{Ball}, W.~H., \& {Gizon}, L. 2014,
  {\mhref{http://doi.org/10.1051/0004-6361/201424325}{\aap}},
  {\href{https://ui.adsabs.harvard.edu/abs/2014A&A...568A.123B}{568}}{\href{https://ui.adsabs.harvard.edu/abs/2014A&A...568A.123B}{,
  A123}}

\bibitem[{{Basu} {et~al.}(2009){Basu}, {Chaplin}, {Elsworth}, {New}, \&
  {Serenelli}}]{basu_fresh_2009}
{Basu}, S., {Chaplin}, W.~J., {Elsworth}, Y., {New}, R., \& {Serenelli}, A.~M.
  2009, {\mhref{http://doi.org/10.1088/0004-637X/699/2/1403}{\apj}},
  {\href{https://ui.adsabs.harvard.edu/abs/2009ApJ...699.1403B}{699}}{\href{https://ui.adsabs.harvard.edu/abs/2009ApJ...699.1403B}{,
  1403}}

\bibitem[{{Bellinger} {et~al.}(2021){Bellinger}, {Basu}, {Hekker},
  {Christensen-Dalsgaard}, \& {Ball}}]{bellinger_asteroseismic_2021}
{Bellinger}, E.~P., {Basu}, S., {Hekker}, S., {Christensen-Dalsgaard}, J., \&
  {Ball}, W.~H. 2021, {\mhref{http://doi.org/10.3847/1538-4357/ac0051}{\apj}},
  {\href{https://ui.adsabs.harvard.edu/abs/2021ApJ...915..100B}{915}}{\href{https://ui.adsabs.harvard.edu/abs/2021ApJ...915..100B}{,
  100}}

\bibitem[{{Christensen-Dalsgaard} \& {Schou}(1988)}]{jcd_differential_1996}
{Christensen-Dalsgaard}, J., \& {Schou}, J. 1988, in ESA Special Publication,
  Vol. 286, Seismology of the Sun and Sun-Like Stars, ed. E.~J. {Rolfe},
  {\href{https://ui.adsabs.harvard.edu/abs/1988ESASP.286..149C}{149--153}}

\bibitem[{{Christensen-Dalsgaard} {et~al.}(1990){Christensen-Dalsgaard},
  {Schou}, \& {Thompson}}]{jcd_comparison_1990}
{Christensen-Dalsgaard}, J., {Schou}, J., \& {Thompson}, M.~J. 1990,
  {\mhref{http://doi.org/10.1093/mnras/242.3.353}{\mnras}},
  {\href{https://ui.adsabs.harvard.edu/abs/1990MNRAS.242..353C}{242}}{\href{https://ui.adsabs.harvard.edu/abs/1990MNRAS.242..353C}{,
  353}}

\bibitem[{{Deheuvels} {et~al.}(2015){Deheuvels}, {Ballot}, {Beck}, {Mosser},
  {{\O}stensen}, {Garc{\'\i}a}, \& {Goupil}}]{deheuvels_seismic_2015}
{Deheuvels}, S., {Ballot}, J., {Beck}, P.~G., {et~al.} 2015,
  {\mhref{http://doi.org/10.1051/0004-6361/201526449}{\aap}},
  {\href{https://ui.adsabs.harvard.edu/abs/2015A&A...580A..96D}{580}}{\href{https://ui.adsabs.harvard.edu/abs/2015A&A...580A..96D}{,
  A96}}

\bibitem[{{Deheuvels} {et~al.}(2017){Deheuvels}, {Ouazzani}, \&
  {Basu}}]{deheuvels_near_2017}
{Deheuvels}, S., {Ouazzani}, R.~M., \& {Basu}, S. 2017,
  {\mhref{http://doi.org/10.1051/0004-6361/201730786}{\aap}},
  {\href{https://ui.adsabs.harvard.edu/abs/2017A&A...605A..75D}{605}}{\href{https://ui.adsabs.harvard.edu/abs/2017A&A...605A..75D}{,
  A75}}

\bibitem[{{Deheuvels} {et~al.}(2012){Deheuvels}, {Garc{\'\i}a}, {Chaplin},
  {Basu}, {Antia}, {Appourchaux}, {Benomar}, {Davies}, {Elsworth}, {Gizon},
  {Goupil}, {Reese}, {Regulo}, {Schou}, {Stahn}, {Casagrande},
  {Christensen-Dalsgaard}, {Fischer}, {Hekker}, {Kjeldsen}, {Mathur}, {Mosser},
  {Pinsonneault}, {Valenti}, {Christiansen}, {Kinemuchi}, \&
  {Mullally}}]{deheuvels_seismic_2012}
{Deheuvels}, S., {Garc{\'\i}a}, R.~A., {Chaplin}, W.~J., {et~al.} 2012,
  {\mhref{http://doi.org/10.1088/0004-637X/756/1/19}{\apj}},
  {\href{https://ui.adsabs.harvard.edu/abs/2012ApJ...756...19D}{756}}{\href{https://ui.adsabs.harvard.edu/abs/2012ApJ...756...19D}{,
  19}}

\bibitem[{{Deheuvels} {et~al.}(2014){Deheuvels}, {Do{\u{g}}an}, {Goupil},
  {Appourchaux}, {Benomar}, {Bruntt}, {Campante}, {Casagrande}, {Ceillier},
  {Davies}, {De Cat}, {Fu}, {Garc{\'\i}a}, {Lobel}, {Mosser}, {Reese},
  {Regulo}, {Schou}, {Stahn}, {Thygesen}, {Yang}, {Chaplin},
  {Christensen-Dalsgaard}, {Eggenberger}, {Gizon}, {Mathis},
  {Molenda-{\.Z}akowicz}, \& {Pinsonneault}}]{deheuvels_seismic_2014}
{Deheuvels}, S., {Do{\u{g}}an}, G., {Goupil}, M.~J., {et~al.} 2014,
  {\mhref{http://doi.org/10.1051/0004-6361/201322779}{\aap}},
  {\href{https://ui.adsabs.harvard.edu/abs/2014A&A...564A..27D}{564}}{\href{https://ui.adsabs.harvard.edu/abs/2014A&A...564A..27D}{,
  A27}}

\bibitem[{{Di Mauro} {et~al.}(2016){Di Mauro}, {Ventura}, {Cardini}, {Stello},
  {Christensen-Dalsgaard}, {Dziembowski}, {Patern{\`o}}, {Beck}, {Bloemen},
  {Davies}, {De Smedt}, {Elsworth}, {Garc{\'\i}a}, {Hekker}, {Mosser}, \&
  {Tkachenko}}]{dimauro_internal_2016}
{Di Mauro}, M.~P., {Ventura}, R., {Cardini}, D., {et~al.} 2016,
  {\mhref{http://doi.org/10.3847/0004-637X/817/1/65}{\apj}},
  {\href{https://ui.adsabs.harvard.edu/abs/2016ApJ...817...65D}{817}}{\href{https://ui.adsabs.harvard.edu/abs/2016ApJ...817...65D}{,
  65}}

\bibitem[{{Fellay} {et~al.}(2021){Fellay}, {Buldgen}, {Eggenberger}, {Khan},
  {Salmon}, {Miglio}, \& {Montalb{\'a}n}}]{fellay_asteroseismology_2021}
{Fellay}, L., {Buldgen}, G., {Eggenberger}, P., {et~al.} 2021,
  {\mhref{http://doi.org/10.1051/0004-6361/202140518}{\aap}},
  {\href{https://ui.adsabs.harvard.edu/abs/2021A&A...654A.133F}{654}}{\href{https://ui.adsabs.harvard.edu/abs/2021A&A...654A.133F}{,
  A133}}

\bibitem[{{Fuller} {et~al.}(2019){Fuller}, {Piro}, \&
  {Jermyn}}]{fuller_slowing_2019}
{Fuller}, J., {Piro}, A.~L., \& {Jermyn}, A.~S. 2019,
  {\mhref{http://doi.org/10.1093/mnras/stz514}{\mnras}},
  {\href{https://ui.adsabs.harvard.edu/abs/2019MNRAS.485.3661F}{485}}{\href{https://ui.adsabs.harvard.edu/abs/2019MNRAS.485.3661F}{,
  3661}}

\bibitem[{{Gehan} {et~al.}(2018){Gehan}, {Mosser}, {Michel}, {Samadi}, \&
  {Kallinger}}]{gehan_core_2018}
{Gehan}, C., {Mosser}, B., {Michel}, E., {Samadi}, R., \& {Kallinger}, T. 2018,
  {\mhref{http://doi.org/10.1051/0004-6361/201832822}{\aap}},
  {\href{https://ui.adsabs.harvard.edu/abs/2018A&A...616A..24G}{616}}{\href{https://ui.adsabs.harvard.edu/abs/2018A&A...616A..24G}{,
  A24}}

\bibitem[{{Gough}(1985)}]{gough_inverting_1985}
{Gough}, D. 1985, {\mhref{http://doi.org/10.1007/BF00158422}{\solphys}},
  {\href{https://ui.adsabs.harvard.edu/abs/1985SoPh..100...65G}{100}}{\href{https://ui.adsabs.harvard.edu/abs/1985SoPh..100...65G}{,
  65}}

\bibitem[{{Gough}(1993)}]{gough_linear_1993}
{Gough}, D.~O. 1993, in Astrophysical Fluid Dynamics - Les Houches 1987, ed.
  J.-P. {Zahn} \& J.~{Zinn-Justin} (Amsterdam: North-Holland),
  {\href{http://adsabs.harvard.edu/abs/1993afd..conf..399G}{399--560}}

\bibitem[{{Gough} \& {Thompson}(1990)}]{gough_rotation_1990}
{Gough}, D.~O., \& {Thompson}, M.~J. 1990,
  {\mhref{http://doi.org/10.1093/mnras/242.1.25}{\mnras}},
  {\href{https://ui.adsabs.harvard.edu/abs/1990MNRAS.242...25G}{242}}{\href{https://ui.adsabs.harvard.edu/abs/1990MNRAS.242...25G}{,
  25}}

\bibitem[{{Goupil} {et~al.}(2013){Goupil}, {Mosser}, {Marques}, {Ouazzani},
  {Belkacem}, {Lebreton}, \& {Samadi}}]{goupil_seismic_2013}
{Goupil}, M.~J., {Mosser}, B., {Marques}, J.~P., {et~al.} 2013,
  {\mhref{http://doi.org/10.1051/0004-6361/201220266}{\aap}},
  {\href{https://ui.adsabs.harvard.edu/abs/2013A&A...549A..75G}{549}}{\href{https://ui.adsabs.harvard.edu/abs/2013A&A...549A..75G}{,
  A75}}

\bibitem[{{Harris} {et~al.}(2020){Harris}, {Millman}, {van der Walt},
  {Gommers}, {Virtanen}, {Cournapeau}, {Wieser}, {Taylor}, {Berg}, {Smith},
  {Kern}, {Picus}, {Hoyer}, {van Kerkwijk}, {Brett}, {Haldane}, {del R{\'\i}o},
  {Wiebe}, {Peterson}, {G{\'e}rard-Marchant}, {Sheppard}, {Reddy}, {Weckesser},
  {Abbasi}, {Gohlke}, \& {Oliphant}}]{numpy}
{Harris}, C.~R., {Millman}, K.~J., {van der Walt}, S.~J., {et~al.} 2020,
  {\mhref{http://doi.org/10.1038/s41586-020-2649-2}{\nat}},
  {\href{https://ui.adsabs.harvard.edu/abs/2020Natur.585..357H}{585}}{\href{https://ui.adsabs.harvard.edu/abs/2020Natur.585..357H}{,
  357}}

\bibitem[{Killingbeck(1977)}]{killingbeck_perturbation_1977}
Killingbeck, J. 1977,
  {\mhref{http://doi.org/10.1088/0034-4885/40/9/001}{Reports on Progress in
  Physics}}, 40, 963

\bibitem[{{Klion} \& {Quataert}(2017)}]{klion_diagnostic_2017}
{Klion}, H., \& {Quataert}, E. 2017,
  {\mhref{http://doi.org/10.1093/mnrasl/slw171}{\mnras}},
  {\href{https://ui.adsabs.harvard.edu/abs/2017MNRAS.464L..16K}{464}}{\href{https://ui.adsabs.harvard.edu/abs/2017MNRAS.464L..16K}{,
  L16}}

\bibitem[{{Kosovichev}(1999)}]{kosovichev_inversion_1999}
{Kosovichev}, A.~G. 1999, Journal of Computational and Applied Mathematics,
  {\href{https://ui.adsabs.harvard.edu/abs/1999JCoAM.109....1K}{109}}{\href{https://ui.adsabs.harvard.edu/abs/1999JCoAM.109....1K}{,
  1}}

\bibitem[{{Landau} \& {Lifshitz}(1965)}]{landau_quantum_1965}
{Landau}, L.~D., \& {Lifshitz}, E.~M. 1965, {Quantum Mechanics. Nonrelativistic
  theory} (Oxford: Pergamon Press)

\bibitem[{{Li} {et~al.}(2023){Li}, {Bedding}, {Stello}, {Huber}, {Hon},
  {Joyce}, {Li}, {Perkins}, {White}, {Zinn}, {Howard}, {Isaacson}, {Hey}, \&
  {Kjeldsen}}]{li_surface_2023}
{Li}, Y., {Bedding}, T.~R., {Stello}, D., {et~al.} 2023,
  {\mhref{http://doi.org/10.1093/mnras/stad1445}{\mnras}},
  {\href{https://ui.adsabs.harvard.edu/abs/2023MNRAS.523..916L}{523}}{\href{https://ui.adsabs.harvard.edu/abs/2023MNRAS.523..916L}{,
  916}}

\bibitem[{{Lynden-Bell} \& {Ostriker}(1967)}]{lyndenbell_stability_1967}
{Lynden-Bell}, D., \& {Ostriker}, J.~P. 1967,
  {\mhref{http://doi.org/10.1093/mnras/136.3.293}{\mnras}},
  {\href{https://ui.adsabs.harvard.edu/abs/1967MNRAS.136..293L}{136}}{\href{https://ui.adsabs.harvard.edu/abs/1967MNRAS.136..293L}{,
  293}}

\bibitem[{{Marchenkov} {et~al.}(2000){Marchenkov}, {Roxburgh}, \&
  {Vorontsov}}]{marchenkov_nonlinear_2000}
{Marchenkov}, K., {Roxburgh}, I., \& {Vorontsov}, S. 2000,
  {\mhref{http://doi.org/10.1046/j.1365-8711.2000.03059.x}{\mnras}},
  {\href{https://ui.adsabs.harvard.edu/abs/2000MNRAS.312...39M}{312}}{\href{https://ui.adsabs.harvard.edu/abs/2000MNRAS.312...39M}{,
  39}}

\bibitem[{{Mosser} {et~al.}(2018){Mosser}, {Gehan}, {Belkacem}, {Samadi},
  {Michel}, \& {Goupil}}]{mosser_period_2018}
{Mosser}, B., {Gehan}, C., {Belkacem}, K., {et~al.} 2018,
  {\mhref{http://doi.org/10.1051/0004-6361/201832777}{\aap}},
  {\href{https://ui.adsabs.harvard.edu/abs/2018A&A...618A.109M}{618}}{\href{https://ui.adsabs.harvard.edu/abs/2018A&A...618A.109M}{,
  A109}}

\bibitem[{{Mosser} {et~al.}(2012){Mosser}, {Elsworth}, {Hekker}, {Huber},
  {Kallinger}, {Mathur}, {Belkacem}, {Goupil}, {Samadi}, {Barban}, {Bedding},
  {Chaplin}, {Garc{\'\i}a}, {Stello}, {De Ridder}, {Middour}, {Morris}, \&
  {Quintana}}]{mosser_power_2012}
{Mosser}, B., {Elsworth}, Y., {Hekker}, S., {et~al.} 2012,
  {\mhref{http://doi.org/10.1051/0004-6361/20111735210.1086/141952}{\aap}},
  {\href{https://ui.adsabs.harvard.edu/abs/2012A&A...537A..30M}{537}}{\href{https://ui.adsabs.harvard.edu/abs/2012A&A...537A..30M}{,
  A30}}

\bibitem[{{Ong} \& {Basu}(2020)}]{ong_semianalytic_2020}
{Ong}, J.~M.~J., \& {Basu}, S. 2020,
  {\mhref{http://doi.org/10.3847/1538-4357/ab9ffb}{\apj}},
  {\href{https://ui.adsabs.harvard.edu/abs/2020ApJ...898..127O}{898}}{\href{https://ui.adsabs.harvard.edu/abs/2020ApJ...898..127O}{,
  127}}

\bibitem[{{Ong} {et~al.}(2021){Ong}, {Basu}, \& {Roxburgh}}]{ong_surface_2021}
{Ong}, J.~M.~J., {Basu}, S., \& {Roxburgh}, I.~W. 2021,
  {\mhref{http://doi.org/10.3847/1538-4357/ac12ca}{\apj}},
  {\href{https://ui.adsabs.harvard.edu/abs/2021ApJ...920....8O}{920}}{\href{https://ui.adsabs.harvard.edu/abs/2021ApJ...920....8O}{,
  8}}

\bibitem[{{Ong} {et~al.}(2022){Ong}, {Bugnet}, \& {Basu}}]{ong_rotation_2022}
{Ong}, J.~M.~J., {Bugnet}, L., \& {Basu}, S. 2022,
  {\mhref{http://doi.org/10.3847/1538-4357/ac97e7}{\apj}},
  {\href{https://ui.adsabs.harvard.edu/abs/2022ApJ...940...18O}{940}}{\href{https://ui.adsabs.harvard.edu/abs/2022ApJ...940...18O}{,
  18}}

\bibitem[{{Ong} \& Gehan(2023)}]{ong_rotation_2023}
{Ong}, J.~M.~J., \& Gehan, C. 2023,
  {\mhref{http://doi.org/10.3847/1538-4357/acbf2f}{\apj}},
  {\href{https://ui.adsabs.harvard.edu/abs/2023arXiv230212402O}{946}}{\href{https://ui.adsabs.harvard.edu/abs/2023arXiv230212402O}{,
  92}}

\bibitem[{{Ouazzani} {et~al.}(2013){Ouazzani}, {Goupil}, {Dupret}, \&
  {Marques}}]{ouazzani_rotation_2013}
{Ouazzani}, R.~M., {Goupil}, M.~J., {Dupret}, M.~A., \& {Marques}, J.~P. 2013,
  {\mhref{http://doi.org/10.1051/0004-6361/201220547}{\aap}},
  {\href{https://ui.adsabs.harvard.edu/abs/2013A&A...554A..80O}{554}}{\href{https://ui.adsabs.harvard.edu/abs/2013A&A...554A..80O}{,
  A80}}

\bibitem[{{Paxton} {et~al.}(2011){Paxton}, {Bildsten}, {Dotter}, {Herwig},
  {Lesaffre}, \& {Timmes}}]{mesa_paper_1}
{Paxton}, B., {Bildsten}, L., {Dotter}, A., {et~al.} 2011,
  {\mhref{http://doi.org/10.1088/0067-0049/192/1/3}{\apjs}},
  {\href{http://adsabs.harvard.edu/abs/2011ApJS..192....3P}{192}}{\href{http://adsabs.harvard.edu/abs/2011ApJS..192....3P}{,
  3}}

\bibitem[{{Paxton} {et~al.}(2013){Paxton}, {Cantiello}, {Arras}, {Bildsten},
  {Brown}, {Dotter}, {Mankovich}, {Montgomery}, {Stello}, {Timmes}, \&
  {Townsend}}]{mesa_paper_2}
{Paxton}, B., {Cantiello}, M., {Arras}, P., {et~al.} 2013,
  {\mhref{http://doi.org/10.1088/0067-0049/208/1/4}{\apjs}},
  {\href{http://adsabs.harvard.edu/abs/2013ApJS..208....4P}{208}}{\href{http://adsabs.harvard.edu/abs/2013ApJS..208....4P}{,
  4}}

\bibitem[{{Paxton} {et~al.}(2015){Paxton}, {Marchant}, {Schwab}, {Bauer},
  {Bildsten}, {Cantiello}, {Dessart}, {Farmer}, {Hu}, {Langer}, {Townsend},
  {Townsley}, \& {Timmes}}]{mesa_paper_3}
{Paxton}, B., {Marchant}, P., {Schwab}, J., {et~al.} 2015,
  {\mhref{http://doi.org/10.1088/0067-0049/220/1/15}{\apjs}},
  {\href{https://ui.adsabs.harvard.edu/abs/2015ApJS..220...15P}{220}}{\href{https://ui.adsabs.harvard.edu/abs/2015ApJS..220...15P}{,
  15}}

\bibitem[{{Paxton} {et~al.}(2018){Paxton}, {Schwab}, {Bauer}, {Bildsten},
  {Blinnikov}, {Duffell}, {Farmer}, {Goldberg}, {Marchant}, {Sorokina},
  {Thoul}, {Townsend}, \& {Timmes}}]{mesa_paper_4}
{Paxton}, B., {Schwab}, J., {Bauer}, E.~B., {et~al.} 2018,
  {\mhref{http://doi.org/10.3847/1538-4365/aaa5a8}{\apjs}},
  {\href{http://adsabs.harvard.edu/abs/2018ApJS..234...34P}{234}}{\href{http://adsabs.harvard.edu/abs/2018ApJS..234...34P}{,
  34}}

\bibitem[{{Paxton} {et~al.}(2019){Paxton}, {Smolec}, {Schwab}, {Gautschy},
  {Bildsten}, {Cantiello}, {Dotter}, {Farmer}, {Goldberg}, {Jermyn}, {Kanbur},
  {Marchant}, {Thoul}, {Townsend}, {Wolf}, {Zhang}, \& {Timmes}}]{mesa_paper_5}
{Paxton}, B., {Smolec}, R., {Schwab}, J., {et~al.} 2019,
  {\mhref{http://doi.org/10.3847/1538-4365/ab2241}{\apjs}},
  {\href{https://ui.adsabs.harvard.edu/abs/2019ApJS..243...10P}{243}}{\href{https://ui.adsabs.harvard.edu/abs/2019ApJS..243...10P}{,
  10}}

\bibitem[{{Pijpers} {et~al.}(2021){Pijpers}, {Di Mauro}, \&
  {Ventura}}]{pijpers_asteroseismogyrometry_2021}
{Pijpers}, F.~P., {Di Mauro}, M.~P., \& {Ventura}, R. 2021,
  {\mhref{http://doi.org/10.1051/0004-6361/202140933}{\aap}},
  {\href{https://ui.adsabs.harvard.edu/abs/2021A&A...656A.151P}{656}}{\href{https://ui.adsabs.harvard.edu/abs/2021A&A...656A.151P}{,
  A151}}

\bibitem[{{Pijpers} \& {Thompson}(1994)}]{pijpers_sola_1994}
{Pijpers}, F.~P., \& {Thompson}, M.~J. 1994, \aap,
  {\href{https://ui.adsabs.harvard.edu/abs/1994A&A...281..231P}{281}}{\href{https://ui.adsabs.harvard.edu/abs/1994A&A...281..231P}{,
  231}}

\bibitem[{{Rabello-Soares} {et~al.}(1999){Rabello-Soares}, {Basu}, \&
  {Christensen-Dalsgaard}}]{rabellosoares_parameters_1999}
{Rabello-Soares}, M.~C., {Basu}, S., \& {Christensen-Dalsgaard}, J. 1999,
  {\mhref{http://doi.org/10.1046/j.1365-8711.1999.02785.x}{\mnras}},
  {\href{https://ui.adsabs.harvard.edu/abs/1999MNRAS.309...35R}{309}}{\href{https://ui.adsabs.harvard.edu/abs/1999MNRAS.309...35R}{,
  35}}

\bibitem[{{Reback} {et~al.}(2021){Reback}, {Mendel}, {McKinney}, {Van Den
  Bossche}, {Augspurger}, {Cloud}, {Hawkins}, {Gfyoung}, {Sinhrks}, {Roeschke},
  {Klein}, {Petersen}, {Tratner}, {She}, {Ayd}, {Hoefler}, {Naveh}, {Garcia},
  {Schendel}, {Hayden}, {Saxton}, {Gorelli}, {Shadrach}, {Jancauskas},
  {McMaster}, {Li}, {Battiston}, {Seabold}, {Attack68}, \& {Dong}}]{pandas}
{Reback}, J., {Mendel}, J.~B., {McKinney}, W., {et~al.} 2021,
  {pandas-dev/pandas: Pandas 1.3.0}, v1.3.0,  Zenodo,
  \dodoi{10.5281/zenodo.3509134}

\bibitem[{{Roxburgh} \& {Vorontsov}(1996)}]{roxburgh_asymptotic_1996}
{Roxburgh}, I.~W., \& {Vorontsov}, S.~V. 1996,
  {\mhref{http://doi.org/10.1093/mnras/278.4.940}{\mnras}},
  {\href{https://ui.adsabs.harvard.edu/abs/1996MNRAS.278..940R}{278}}{\href{https://ui.adsabs.harvard.edu/abs/1996MNRAS.278..940R}{,
  940}}

\bibitem[{{Schr{\"o}dinger}(1926)}]{schrodinger_quantisierung_1926}
{Schr{\"o}dinger}, E. 1926,
  {\mhref{http://doi.org/10.1002/andp.19263851302}{Annalen der Physik}},
  {\href{https://ui.adsabs.harvard.edu/abs/1926AnP...385..437S}{385}}{\href{https://ui.adsabs.harvard.edu/abs/1926AnP...385..437S}{,
  437}}

\bibitem[{{Schunker} {et~al.}(2016{\natexlab{a}}){Schunker}, {Schou}, \&
  {Ball}}]{schunker_inversion_2016a}
{Schunker}, H., {Schou}, J., \& {Ball}, W.~H. 2016{\natexlab{a}},
  {\mhref{http://doi.org/10.1051/0004-6361/201525937}{\aap}},
  {\href{https://ui.adsabs.harvard.edu/abs/2016A&A...586A..24S}{586}}{\href{https://ui.adsabs.harvard.edu/abs/2016A&A...586A..24S}{,
  A24}}

\bibitem[{{Schunker} {et~al.}(2016{\natexlab{b}}){Schunker}, {Schou}, {Ball},
  {Nielsen}, \& {Gizon}}]{schunker_inversion_2016b}
{Schunker}, H., {Schou}, J., {Ball}, W.~H., {Nielsen}, M.~B., \& {Gizon}, L.
  2016{\natexlab{b}},
  {\mhref{http://doi.org/10.1051/0004-6361/201527485}{\aap}},
  {\href{https://ui.adsabs.harvard.edu/abs/2016A&A...586A..79S}{586}}{\href{https://ui.adsabs.harvard.edu/abs/2016A&A...586A..79S}{,
  A79}}

\bibitem[{{Shibahashi}(1979)}]{shibahashi_modal_1979}
{Shibahashi}, H. 1979, \pasj,
  {\href{https://ui.adsabs.harvard.edu/abs/1979PASJ...31...87S}{31}}{\href{https://ui.adsabs.harvard.edu/abs/1979PASJ...31...87S}{,
  87}}

\bibitem[{{Tayar} {et~al.}(2022){Tayar}, {Moyano}, {Soares-Furtado}, {Escorza},
  {Joyce}, {Martell}, {Garc{\'\i}a}, {Breton}, {Mathis}, {Mathur}, {Delsanti},
  {Kiefer}, {Reffert}, {Bowman}, {Van Reeth}, {Shetye}, {Gehan}, \&
  {Grunblatt}}]{tayar_spinning_2022}
{Tayar}, J., {Moyano}, F.~D., {Soares-Furtado}, M., {et~al.} 2022,
  {\mhref{http://doi.org/10.3847/1538-4357/ac9312}{\apj}},
  {\href{https://ui.adsabs.harvard.edu/abs/2022ApJ...940...23T}{940}}{\href{https://ui.adsabs.harvard.edu/abs/2022ApJ...940...23T}{,
  23}}

\bibitem[{{Townsend} \& {Teitler}(2013)}]{townsend_gyre_2013}
{Townsend}, R.~H.~D., \& {Teitler}, S.~A. 2013,
  {\mhref{http://doi.org/10.1093/mnras/stt1533}{\mnras}},
  {\href{http://adsabs.harvard.edu/abs/2013MNRAS.435.3406T}{435}}{\href{http://adsabs.harvard.edu/abs/2013MNRAS.435.3406T}{,
  3406}}

\bibitem[{{Triana} {et~al.}(2017){Triana}, {Corsaro}, {De Ridder}, {Bonanno},
  {P{\'e}rez Hern{\'a}ndez}, \& {Garc{\'\i}a}}]{triana_internal_2017}
{Triana}, S.~A., {Corsaro}, E., {De Ridder}, J., {et~al.} 2017,
  {\mhref{http://doi.org/10.1051/0004-6361/201629186}{\aap}},
  {\href{https://ui.adsabs.harvard.edu/abs/2017A&A...602A..62T}{602}}{\href{https://ui.adsabs.harvard.edu/abs/2017A&A...602A..62T}{,
  A62}}

\bibitem[{{Unno} {et~al.}(1989){Unno}, {Osaki}, {Ando}, {Saio}, \&
  {Shibahashi}}]{unno_nonradial_1989}
{Unno}, W., {Osaki}, Y., {Ando}, H., {Saio}, H., \& {Shibahashi}, H. 1989,
  {Nonradial oscillations of stars} (Tokyo: University of Tokyo Press)

\bibitem[{{Vanlaer} {et~al.}(2023){Vanlaer}, {Aerts}, {Bellinger}, \&
  {Christensen-Dalsgaard}}]{vanlaer_feasibility_2023}
{Vanlaer}, V., {Aerts}, C., {Bellinger}, E.~P., \& {Christensen-Dalsgaard}, J.
  2023, {\mhref{http://doi.org/10.1051/0004-6361/202245597}{\aap}},
  {\href{https://ui.adsabs.harvard.edu/abs/2023A&A...675A..17V}{675}}{\href{https://ui.adsabs.harvard.edu/abs/2023A&A...675A..17V}{,
  A17}}

\bibitem[{{Virtanen} {et~al.}(2020){Virtanen}, {Gommers}, {Oliphant},
  {Haberland}, {Reddy}, {Cournapeau}, {Burovski}, {Peterson}, {Weckesser},
  {Bright}, {van der Walt}, {Brett}, {Wilson}, {Millman}, {Mayorov}, {Nelson},
  {Jones}, {Kern}, {Larson}, {Carey}, {Polat}, {Feng}, {Moore}, {VanderPlas},
  {Laxalde}, {Perktold}, {Cimrman}, {Henriksen}, {Quintero}, {Harris},
  {Archibald}, {Ribeiro}, {Pedregosa}, {van Mulbregt}, \& {SciPy 1. 0
  Contributors}}]{scipy}
{Virtanen}, P., {Gommers}, R., {Oliphant}, T.~E., {et~al.} 2020,
  {\mhref{http://doi.org/10.1038/s41592-019-0686-2}{NatMe}},
  {\href{https://ui.adsabs.harvard.edu/abs/2020NatMe..17..261V}{17}}{\href{https://ui.adsabs.harvard.edu/abs/2020NatMe..17..261V}{,
  261}}

\bibitem[{{Wilson} {et~al.}(2023){Wilson}, {Casey}, {Mandel}, {Ball},
  {Bellinger}, \& {Davies}}]{wilson_constraining_2023}
{Wilson}, T.~A., {Casey}, A.~R., {Mandel}, I., {et~al.} 2023,
  {\mhref{http://doi.org/10.1093/mnras/stad771}{\mnras}},
  {\href{https://ui.adsabs.harvard.edu/abs/2023MNRAS.521.4122W}{521}}{\href{https://ui.adsabs.harvard.edu/abs/2023MNRAS.521.4122W}{,
  4122}}

\end{thebibliography}

\end{document}